# Mutual and self-inductance in planarized multilayered superconductor integrated circuits: Microstrips, striplines, bends, meanders, ground plane perforations

Sergey K. Tolpygo, Evan B. Golden, Terence J. Weir, and Vladimir Bolkhovsky

Lincoln Laboratory, Massachusetts Institute of Technology, Lexington, MA, United States of America

E-mail: sergey.tolpygo@ll.mit.edu

**Abstract**

Data are presented on mutual and self-inductance of various inductors used in multilayered superconductor integrated circuits: microstrips and striplines with widths of signal traces from 250 nm to a few micrometers, located on the same circuit layer at various distances from each other (from 250 nm to a few micrometers) and/or on different layers spaced vertically; effect of long slits in the ground plane(s) along the inductors on their mutual inductance; inductance of right-angled bends; inductance of meanders. Simple analytical expressions for mutual and self-inductance of the basic inductors are given, describing experimental data with accuracy better than 2% in a very wide range of parameters. They can be used for superconductor integrated circuit design and calibration of numerical inductance extractors. Measurements were done using circuits fabricated in fully planarized fabrication processes with eight niobium layers and Nb/Al-AlO$_x$/Nb Josephson junctions, known as the SFQ5ee and SC1 processes developed at MIT Lincoln Laboratory for superconductor electronics. Mutual inductance decreases exponentially with distance between striplines and as a second power of the distance between microstrips, strongly depends on magnetic field penetration depths in superconducting ground planes, whereas superconducting properties of the signal traces are practically immaterial. Weak dependence of mutual inductance on the linewidth of superconducting wires indicates that area of superconducting flux transformers - the essential component of all digital circuits using ac power, superconducting qubits, and sensor arrays - scales poorly with the linewidth, putting a predictable upper limit on the integration scale of such circuits.

## 1. Introduction

Recent advances in fabrication technology have enabled fabrication of niobium-based superconductor integrated circuits with feature sizes down to 150 nm, increased circuit density above $10^7$ Nb/Al-AlO$_x$/Nb Josephson junctions (JJs) per square centimeter [1, 2], about a 10x increase from the previous level [3], and increased the total number of JJs and inductors in functional circuits to about ten millions. Increasing the scale of integration introduces new challenges − closely spaced superconducting wires become mutually coupled and can no longer be considered as isolated inductors. Accurate account for mutual inductance of features comprising superconductor logic and memory cells, qubits, sensors, etc. is essential for designing functioning integrated circuits because mutual coupling may completely alter their performance or reduce margins of operation, rendering circuits nonoperational in extreme cases. For instance, a successful demonstration of a compact Josephson memory cell containing four Josephson junctions and six inductors required taking into account mutual coupling of five pairs of inductors in order to achieve a functional density of 1 Mbit cm$^{-2}$ [4]. On the other hand, with progress in miniaturization, inductors (wires) in integrated circuits are becoming nearly thin filaments with about square cross section and dimensions comparable to magnetic field penetration depth $\lambda$ in the superconducting material of the circuit layers (currently mainly niobium). In many cases, this allows to obtain very simple and accurate analytical expressions for self- and mutual inductance of basic components such as microstrip and stripline inductors, meanders, etc. They can be used in circuit design and for verification of computer-assisted inductance extraction tools.

There are cases when strong mutual coupling is required, e.g. for making compact superconducting flux transformers. Transformers are an absolutely essential component of all superconductor logic and memory circuits utilizing alternating current (AC) powering and clocking schemes such as Quantum Flux Parametron (QFP) [5, 6] and its adiabatic (AQFP) version [7, 8], and Reciprocal Quantum Logic (RQL) [9] circuits. In them, mutual inductors (transformers) are used to deliver multiphase ac clock power, provide phase shifting and signal inverting (NOT) functions inside logic cells as well as to provide coupling between the cells. In a recently proposed SFQ biasing of SFQ circuits [10, 11] superconducting transformers are also used in AC/SFQ converters delivering power from a single-phase ac power line to SFQ circuitry in the form of single flux quanta. Superconducting transformers are also used in all superconductor quantum circuits as parts of qubits and couplers; see, e.g. [12 - 14]. In superconducting quantum interference devices (SQUIDs), superconducting quantum interference filters (SQIFs), transition edge sensors, and other types of superconducting sensors and arrays of sensors, transformers are used to couple signals in and out of the sensors to readout circuitry or to other sensors. In the original rapid single flux quantum (RSFQ) technology [15] and its energy efficient biasing version ERSFQ [16], inductive coupling is used to transfer clock and data between serially biased parts of the circuits; see [17, 18] and references therein. Negative mutual inductance is used to create the so-called nSQUIDs − SQUIDs with negative inductance between their arms − which were used as building blocks of superconductor reversible circuits [19, 20].

As will be discussed in more detail elsewhere, superconducting flux transformers (mutual inductors) are very poorly scalable circuit component, i.e. they have a very weak dependence of area on the linewidth of superconducting traces. This makes superconductor electronics with its extremely wide-spread use of and reliance on flux transformers a poorly scalable technology in general [21].

A large amount of work was done to study inductance of various structures typically encountered in superconductor integrated circuits fabricated with relatively large, about 1 μm and larger, feature sizes [22−26]. We reported on inductance measurements of circuit features with linewidths down to 250 nm [27, 28] and more recently extended the fabrication and measurements down to 120 nm [29]. The circuits used were fabricated in the most advanced processes developed for superconductor electronics at MIT Lincoln Laboratory (MIT LL): the SFQ5ee process [30, 31], the 250-nm SC1 process [32], and the 150-nm SC2 process [1]. Unfortunately, there are basically no data on mutual inductance of submicron features at submicron spacings allowed in the advanced processes. It is also not clear how well the existing inductance extractors and simulators handle mutual inductance at these dimensions.

Because superconductor integrated circuits contain a large number of inductors, the size of inductance matrix for numerical inductance extraction is typically very large, and finding its elements is extremely computation intensive. Therefore, knowing mutual inductance of various circuit components and its dependence on the distance between them allows circuit designers to properly truncate the inductance matrix by neglecting certain elements due to their smallness.

In order to address these issues and enable circuit design into the advanced nodes of fabrication processes developed at MIT LL, in this work we present mutual inductance data for the main types of inductors used in modern superconductor integrated circuits. We derive simple analytical expressions that accurately describe mutual and self-inductance of superconducting microstrip and stripline inductors in a wide range of parameters, typically with better than 2% accuracy. These expressions can be used in conjunction with superconductor circuit layer tools for computer assisted circuit design and for calibration of numerical inductance extractors. We evaluate a few examples were mutual inductance is essential for determining correctly total inductance of composite inductors comprised multiple primitive inductors, e.g., meanders. We compare results of the measurements with numerical simulations using commercial inductance extraction software: a transmission line extractor wxLC and a full three-dimensional extractor InductEx$^@$ 6.0.

## 2. Fabrication and measurements of test circuits

For measuring self- and mutual inductance, in a course of several years, we fabricated a large number of 5 mm x 5 mm test circuits in all main fabrication processes on 200-mm wafers, developed at MIT LL for superconductor electronics: the standard 8-Nb-layer planarized process SFQ5ee with the nominal minimum linewidth of 350 nm [30, 31]; a more recent 250-nm SC1 process [32]; and a 150-nm SC2 process [1]. Our goal was to analyze and provide data for all of them since they are used by various government programs and by many circuit design teams in industry and academia.

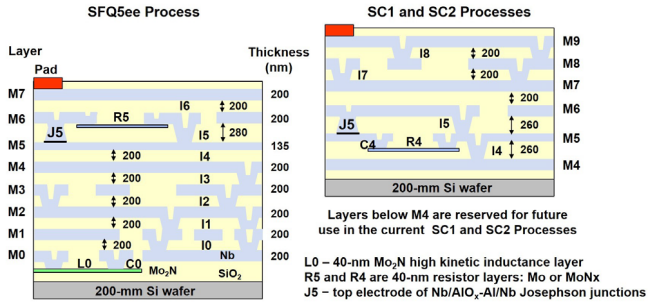

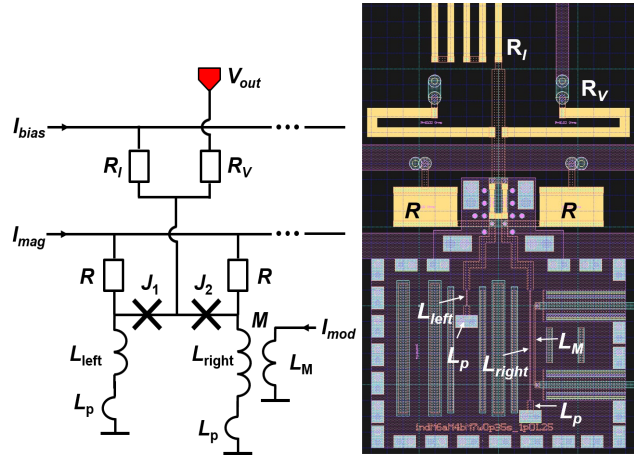

Fig. 1. Cross sections of the SFQ5ee process, the SC1 and the SC2 processes developed at MIT Lincoln Laboratory for superconductor electronics. All layers of SiO$_2$ interlayer dielectric are planarized by chemical meachnical polishing and are 200 nm thick, except layer I5 in the SFQ5ee process, where it is 280 nm, and layers I4 and I5 in the SC1/SC2 processes where they are 260 nm. All Nb layers are 200 nm, except for the layer M5 (base electrode of Josephson junction, which is 135 nm after anodization of Al barrier layer. We use the same notion, e.g., I4 for the dielectric layer between metals M4 and M5 and via to the layer M4. All via are formed by etching contact holes in the interlayer dielectric and filling them with Nb of the next wiring layer. The recommended minimum linewidth of Nb inductors is, respectively, 350, 250, and 150 nm in the SFQ5ee, SC1 and SC2 processes; minimum spacing is 250 nm. For more details, see [1, 30 - 32].

A schematic cross section of the processes used is shown in Fig. 1. The difference between the SC1/SC2 and the SFQ5ee processes is in the location of the Josephson junction layer, J5 in the process stack − it is the fifth Nb layer in the SFQ5ee and the second in the SC1 process. For design compatibility, we preserved the same layer notations in both processes, starting Nb layer count from M0 in the SFQ5ee and from M4 in the SC1. We use the same notations for the dielectric layers, e.g., I4, and for the features patterned in these layers, e.g., I4 vias formed after etched contact holes in I4 dielectric are filled with Nb of the next metal layer, M5.

Interlayer dielectric in all processes is planarized using chemical mechanical polishing (CMP). As a result, each Nb layer is deposited on a smooth planar surface, allowing to pattern deep sub-micrometer features using a deep-UV photolithography. The minimum Nb line spacing, $s$ achievable in the processes is limited by gap fill requirements to 250 nm. The minimum linewidth, $w$ we used for inductors in the SFQ5ee and SC1 processes was 250 nm, and 120 nm in the SC2 node [1] with 193-nm photolithography on a few critical layers. Target (nominal) thicknesses of all layers in the SFQ5ee and the SC1/SC2 processes used in this work are given in Fig. 1.

*A. Self- and Mutual Inductance Measurements*

Inductance and mutual inductance measurements of various features were done using a SQUID-based method [33] and an integrated circuit developed in [27] for a parasitic-free extraction of self- and mutual inductances using a differential method. The typical circuit diagram and a fragment of the circuit layout are shown in Fig. 2. Inductors $L_{left}$ and $L_{right}$ of the same type form two arms of a dc SQUID. Equal currents were fed into both arms from the common magnetic bias rail, $I_{mag}$, using two identical thin-film resistors, $R$, 10 μm long and 20 μm wide. Large dimensions of the resistors and low resistor width standard deviation (1σ < 50 nm) provided equal resistances with better than 1% accuracy. Inductors $L_{left}$ and $L_{right}$ were placed 20 μm apart so their mutual coupling could be completely neglected. They differed only by the length, being identical in all other respects.

Fig. 2. Circuit diagram (left) and layout (right) of one unit of inductance measurement circuit, after [27]. The circuit comprises $N$ such units, $N$ SQUIDs, each formed by identical JJs $J_1$ and $J_2$, biased from a common bias rail $I_{bias}$ via bias resistors $R_I$. SQUID output voltage, $V_{out}$ is measured as a function of magnetic bias current $I_{mag}$ which is fed into both arms of each SQUID via identical resistors, $R$ dividing the current equally between $2N$ inductors. The period $\Delta I_{mag}$ of SQUID voltage modulation $V_{out}(I_{mag})$ is used to calculated the difference of the inductances of the right $L_{right}$ and left $L_{left}$ arms of the SQUID from (1). If the arms, e.g., stripline inductors, differ only by the length, this differential method eliminates parasitic contributions, $L_p$ associated with the inductors' connections to the ground and to the junctions $J_1$ and $J_2$.

From the period of SQUID modulation, $\Delta I_{mag}$, we extracted the difference of inductances of the SQUID arms

$$L = 2N\Phi_0/\Delta I_{mag}, \qquad (1)$$

where $L = |L_{right} - L_{left}|$ and $N$ is the number of SQUIDs connected to the common magnetic bias $I_{mag}$. This differential method allowed us to eliminate parasitic inductance, $L_p$ associated with connection of the inductors under test to the bare SQUID on one end and to the ground plane(s) on the other end. If $\Delta I_{mag}$ is measured with 1% accuracy and the resistors are identical within 1%, the accuracy of this method can be estimated as 1.4%.

Similarly, a mutual inductor, $L_M$ was placed parallel to the inductor $L_{right}$, and a modulation current $I_{mod}$ was fed into it from a separate current source, Fig. 2. Mutual inductance, $M$ of the two inductors determines the period of the SQUID modulation, $\Delta I_{mod}$. The circuits contained 24 dc-SQUID-based test structures per chip, grouped with $N = 6$, and allowed to extract self-inductances of 24 inductors and mutual inductances of 24 pairs of inductors.

In all cases, the differential length, *l* of the individual inductors and the mutual running length of the parallel signal conductors in coupled inductors, $l_M$ was made much larger than the inductor width, *w*, and the spacing, *s*, between the coupled inductors, typically $l \sim 100w$, $100s$. For instance, for all inductors with linewidth $w \leq 0.25$ μm, we used $l = l_M = 30$ μm, and longer for larger linewidths. This allowed us to neglect edge effects of narrow, 0.25-μm, perpendicular wires feeding current into $L_M$. It is convenient to characterize long and uniform inductors by the self- and mutual inductance per unit length, $L_l = L/l$ and $M_l = M/l_M$, and compare the measurements with numerical simulators based on an infinite line approximation. For comparison with a full 3D inductance extractor, the actual layout of inductors in GDSII format was used.

The maximum magnetic modulation current which could be supplied to each SQUID was limited by the critical current of the superconducting wires, $I_c$ which was about 20 mA for the narrow lines used. This sets the lower limit on the mutual and self-inductance which could be measured at about 0.1 pH or at 4 fH/μm for the typical lengths of inductors used.

Hereafter, we use notations M#aM#, e.g., M6aM4, for microstrips, where the first M# indicates the signal conductor layer (e.g., M6) placed above the ground plane (GP) indicated by the second M#, e.g., M4. Inverted microstrips are labeled as M#bM#, e.g., M6bM7 indicates the signal conductor on layer M6 placed below M7 ground (sky) plane. Similarly, various striplines are referred to by indicating the signal conductor first and then the two ground planes, e.g., M6aM4bM7 refers to a stripline with signal conductor formed on the layer M6 placed above M4 and below M7 ground planes. In all stripline cases, the two ground planes were connected together around their edges about 20 μm far away from the inductors in Fig. 2 and also at their grounded ends, using multiple superconducting vias. This created a box-like shielding configuration closely emulating a configuration with two infinite superconducting ground planes.

## 3. Background and inductance calculation methods

Development of methods for calculating self- and mutual inductance of normal conductors of various shapes started with Maxwell who developed the method of geometric mean distances (GMD) [34]. All analytical formulas developed by 1946 were summarized in the classical books by Grover [35-37]. Many simplified expression can also be found in [38-41]. More recent developments for integrated circuits with normal conductors were reviewed by Ruehli [42], who also justified the use of the so-called partial inductances. Below, we briefly review the main definitions and formulas with application to superconductors.

For any superconducting loop, the loop self-inductance can be defined as a coefficient in the expression relating the total energy of the loop, *U* and electric current in the loop, *I*

$$U = U_m + U_k = \tfrac{1}{2}L_m I^2 + \tfrac{1}{2}L_k I^2 = \tfrac{1}{2}LI^2. \quad (1)$$

Here $U_m$ is the energy of magnetic field around and inside the conductor, and $U_k$ is the kinetic energy of the current carriers inside the superconductor. The total inductance *L* is the sum of magnetic, or geometrical, inductance $L_m$ and kinetic inductance $L_k$, where

$$L_k = \frac{\mu_0 m}{2n_s e^2} \frac{\iiint j_s^2 dV}{(\iint j_s dS)^2}, \quad (2)$$

and *m*, *e*, and $n_s$ are electron mass, charge, and Cooper pair number density, respectively, $j_s$ is the superconducting current density, and $\mu_0 = 4\pi \cdot 10^{-7}$ H/m is vacuum permeability. The integral in the numerator of (2) is over the superconductor volume; the integral for the total current in denominator is over the superconductor cross section area, *S*. For a superconductor with uniform current distribution and constant cross section, (2) reduces to the well-known expression

$$L_k = \mu_0(m/2n_s e^2)\, l/S, \quad (3)$$

where *l* is the superconductor length. The quantity in parenthesis defines the London penetration depth of magnetic field into a superconductor $\lambda_L = (m/2n_s e^2)^{1/2}$ [43].

In general, magnetic inductance is a loop property and cannot be defined without specifying the loop, e.g., the return current path, whereas kinetic inductance is a materials property and can be defined for any piece of superconductor. A concept of partial inductances, i.e. inductances of the segments of the loop can also be justified for superconducting loops similar to [42] with the caveat that values of these partial inductances may change if they are inserted into a different loop.

A textbook definition of self-inductance was introduced long before the discovery of superconductivity, see [34]. It relates magnetic flux threading the loop $\Phi_m$ and the loop current

$$\Phi_m = L_m I \quad (4)$$

This definition applies only to the magnetic part of inductance because kinetic inductance *does not create magnetic flux*. Kinetic inductance creates an additional difference in the phase of the superconducting wave function $\Psi = \Psi_0 e^{i\theta}$ between the ends of the loop but does not create magnetic flux. This follows from the quantum-mechanical expression for the current density in superconductors, i.e., the generalized second London equation

$$\vec{J}_s = \frac{m}{2n_s e^2}\left(\frac{\Phi_0}{2\pi}\nabla\theta - \vec{A}\right), \quad (5)$$

where $\Phi_0 = 2e/h$ is the flux quantum and $\vec{A}$ is vector potential; see, e.g. [44]. Taking a path integral inside the superconductor along the current direction from the beginning of the loop at any point 1 on its cross section to any point 2 at the loop end results in

$$\frac{m}{2n_s e^2}\int \vec{J}_s d\vec{l} + \Phi_m = \frac{\Phi_0}{2\pi}(\theta_2 - \theta_1), \quad (6)$$

where $\Phi_m$ is the magnetic flux threading the path. For a closed path, $\theta_2 - \theta_1 = 2\pi n$, where *n* is integer, and the quantity on the left hand side of (6) known as fluxoid becomes quantized. Fluxoid quantization turns into a familiar flux quantization $\Phi_m = n\Phi_0$ if there are no currents in the bulk of the superconducting loop.

Using (2) and (4), we can present (6) as

$$(L_k + L_m)I = LI = \frac{\Phi_0}{2\pi}(\theta_2 - \theta_1). \quad (7)$$

The latter expression is used in superconductor circuit simulation tools such as PSCAN [45, 46] and PSCAN2 [47] to define an inductor as a circuit elements producing a certain phase drop on it per unit current flowing in it. For a closed superconducting loop, eq. (7) gives a generalized quantization condition

$$L_k I + \Phi_m = n\Phi_0. \quad (8)$$

For two inductors (loops) with self-inductances $L_1$ and $L_2$ and current $I_1$ and $I_2$, respectively, a part of magnetic flux created by inductor $L_1$ may leak into inductor $L_2$, adding constructively or destructively to its self-induced flux, and vice versa. Mutual inductance is defined using relationships

$$\Phi_1 = L_1 I_1 + M_{12} I_2 \quad (9a)$$

$$\Phi_2 = L_2 I_2 + M_{21} I_1 \quad (9b)$$

where $\Phi_1$ and $\Phi_2$ are the total magnetic fluxes threading each loop and $M_{12} = M_{21} = M$ is the mutual inductance. Inductors are termed adding inductors, $M > 0$, if the self-flux and the induced flux add constructively, and opposing inductors, $M < 0$, if the induced flux has the opposite sign to the self-flux due to an opposite winding of inductors, or opposite current direction in them. With this definition, a loop self-inductance is mutual inductance of the loop with itself.

Similarly, for $N$ inductors, an inductance matrix is defined as

$$\begin{pmatrix} \Phi_1 \\ \vdots \\ \Phi_N \end{pmatrix} = \begin{pmatrix} L_{11} & \cdots & L_{1N} \\ \vdots & \ddots & \vdots \\ L_{N1} & \cdots & L_{NN} \end{pmatrix} \begin{pmatrix} I_1 \\ \vdots \\ I_N \end{pmatrix}, \quad (10)$$

where all diagonal elements are self-inductances $L_{ii} \equiv L_i$ and all off-diagonal, $i \neq j$, components are mutual inductances $L_{ij} = L_{ji} \equiv M_{ij} = M_{ji}$. Typical logic cells of superconductor integrated circuits may contain a few dozens of inductors.

Note, that definitions (9), (10) include total self-inductance $L = L_m + L_k$. This presents no problem because kinetic inductance does not create magnetic flux. However, since long before superconductors and kinetic inductors, it has been a custom to express mutual inductance as

$$M_{12} = \kappa (L_{1m} L_{2m})^{1/2}, \quad (11)$$

where both inductances are purely magnetic, and $\kappa$ is a coupling constant which depends only on the geometry of the inductors. This may create some confusion because superconductor inductance extraction tools simulate (extract) the full inductance and separating it into two parts is not possible. Hence, if the coupling constant in the software is calculated using $\kappa = M/(L_1 L_2)^{1/2}$, it becomes dependent on the value of kinetic inductances and is no longer a purely geometrical factor. For instance, $\kappa$ would decrease with increasing kinetic inductance, e.g., due to decreasing inductors cross sections or increasing the value of London penetration depth, whereas the actual magnetic coupling in (10) may remain constant.

It is well known that for a serial connection and parallel connection of two inductors, $L_1$ and $L_2$, the total inductance is, respectively,

$$L = L_1 + L_2 + 2M \quad (1a)$$

$$L = \frac{L_1 L_2 - M^2}{L_1 + L_2 - 2M}, \quad (1b)$$

and depends on the value and sign of the mutual inductance, $M$. In deriving these and other expressions for various connections of inductors, e.g. a star (T-connection), triangle, etc., it is assumed that connecting several inductors together does not alter their inductance values. This assumption may not necessarily be true for superconducting circuits because connecting various superconducting thin-film shapes forming inductors between multiple layers usually alters current distribution in them and in superconducting ground planes carrying return currents. Therefore, modeling superconducting inductors in general requires numerical methods and software allowing for inductance and mutual inductance extraction from actual layouts.

Most methods of computing self- and mutual inductance are based on dividing an inductor into filaments – straight pieces of wires with length $\vec{dl}$ and of small cross-section so the current $I$ in them can be treated as uniform. Magnetic field $d\vec{B}(\vec{r})$ created by each filament at a point $\vec{r}$ away from the filmanet is found using Biot-Savart-Laplace law

$$d\vec{B} = \frac{\mu_0}{4\pi} \frac{I \vec{dl} \times \vec{r}}{|\vec{r}|^3}. \quad (12)$$

Then, magnetic flux through any loop is calucalted as a sum (integral) of fluxes $d\Phi = (d\vec{B} \cdot d\vec{S})$ induced by all filaments. For a uniform current distribution, currents in all filaments are the same and equal to the total current divided into the number of filaments in the inductor cross section. For a nonuniform disribution, Maxwell's and London equations need to be solved to find currents in the filaments self-consistently.

Numerical methods and inductance extraction software for superconductors have been subjects of active development, initially for two-dimensional (2-D) and later for 3-D structures; see [48-60] and references therein, in addition to the enormous amount of publications on normal-metal structures. Currently, the most advanced 3-D superconductor inductance extractor is InductEx [60] based on the original FastHenry approach [51, 52] with rectangular filaments (rectangular mesh) and recently extended to include tetrahedral volume elements (triangular mesh) [61]. Very accurate numerical methods were developed by M. Khapaev [54-58], which are based on minimization of the total energy functional and use direct boundary element method to treat conductors of arbitrary cross section. Inductance extraction software developed by Khapaev is known as wxLL for extracting only self- and mutual inductance and as wxLC for extracting inductance, capacitance, wave impedance, and propagation speed in a system of conductors which are uniform in one direction. This software, however, lacks functionality of InductEx and its ability to work with 3D circuit layouts.

In the simplest case of an infinitely long, thin filament (wire) of circular cross section with current $I$, the flux per unit length through any surface bordered by two lines parallel to the axis of the wire at distances $R_1$ and $R_2$ is

$$\Phi_l = \frac{\Phi}{l} = \frac{\mu\mu_0 I}{2\pi} \ln \frac{R_2}{R_1}. \qquad (13)$$

This follows from the Ampere's circuital law for an infinitely long cylindrical wire $B(R) = \frac{\mu\mu_0 I}{2\pi R}$ and Gauss's law, where $R$ is the distance from the wire axis and $\mu$ is the relative magnetic permiability of the space around the wire, which hereafter will be always assumed equal 1.

Equation (13) can be used to find self- and mutual inductances per unit length $L_l = \Phi_l/I$ of uniform conductors with length $l$ much larger than the typical cross-section size when edge effects can be neglected. Wires with more complex cross sections are divided into a large number, $N$ of filaments, each carrying current $I/N$. Then, fluxes induced by each filament and given by (13) need to be summed up. This involves calculating a sum $N^{-1}\sum_i \ln R_i = \ln(R_{gmd})$, where $R_i$ are the distances from each point (filament) of the cross section to the observation point and $R_{gmd}$ is the geometric mean distance from all points of the cross section to the observation point. Maxwell [34] showed that mutual inductance of two infinitely long wires with arbirary (but uniform) cross sections and uniform current distribution can be calculated as mutual inductance of two thin circular filaments placed at a distance $R_{gmd}$ equal to the GMD of all points of one cross section to all points of the other cross section. Since self-inductance is mutual inductance of the wire with itself, it can be calcualted, following Maxwell, as the mutual inductance of two filaments placed at a distance equal to the GMD of all pairs of points of the cross secton, $r_{gmd}$, i.e. the GMD of the cross section to itself.

## 4. Theoretical, numerical, and experimental data for superconducting microstrips

### 4.1 Self- and mutual inductance of superconducting microstrips

*4.1.1 Cylindrical wires.* Consider inductance of an infinitely long superconducting wire of circular cross section placed parallel to a superconducting ground plane with penetration depth $\lambda$, see Fig. 3. The wire radius is $r_1$, and distance from the ground plane to the center of the wire is $h_1$. Due to the boundary conditions on superconducting plane, magnetic field created by the wire with current $+I$ is the same as if the wire image carrying an opposite current $-I$ is placed below the ground plane at a distance $h_1 + \lambda$ from the mirror reflection plane located a distance $\lambda$ below the top surface of the ground plane; see Fig. 3. This differs only slightly from the standard image theory in electrostatics. The distance between the axes of the wire and its image is $2(h_1 + \lambda)$. Using (13), we find the total flux $\Phi_l$ induced per unit length by the current $+I$ and its image $-I$ through a loop between the surface of the wire and the reflection plane and the wire external self-inductance per unit length as $L_l = \Phi_l/I$

$$L_l = \frac{\mu_0}{2\pi}\left(\ln\frac{h_1+\lambda}{r_1} + \ln\frac{2(h_1+\lambda)}{h_1+\lambda}\right). \qquad (14)$$

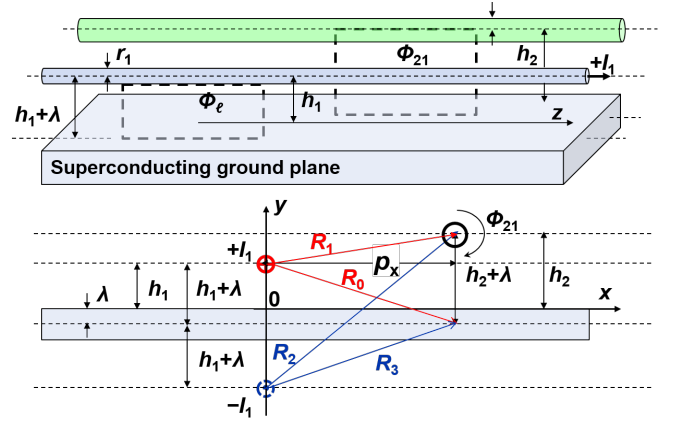

Fig. 3. Two parallel cylindrical wires running in *z*-direction placed above an infinite superconducting ground plane with magnetic field penetration depth $\lambda$ at distances (from their axes), respectively, $h_1$ and $h_2$. Electric current $+I$ in the wire #1 creates an image current $-I$ located symmetrically under the mirror reflection plane at $y = -\lambda$ below the top surface of the superconducting plane as shown in the *x-y* cross section. Both the real and the image currents induce magnetic flux $\Phi_l$ per unit length of the loop formed by wire #1 and the ground plane as well as flux $\Phi_{12}$ per unit length into the second inductor formed by wire #2 and the ground plane.

The first and the second terms in (14) are due, respectively, to the fluxes induced by the wire current and by its image, giving

$$L_l = \frac{\mu_0}{2\pi}\ln\frac{2(h_1+\lambda)}{r_1}. \quad (15)$$

If $r_1 \lesssim \lambda_1$ magnetic field completely penetrates the wire and current in the wire can be consider uniformly distributed; here $\lambda_1$ is magnetic field penetration depth of the wire material which, in general, may differ from the ground plane. Magnetic field inside the wire gives an additional radius-independent contribution $\frac{1}{2}\frac{\mu_0}{4\pi}$ to the magnetic inductance per unit length. Kinetic inductance of carriers in the wire is $\mu_0\lambda_1^2/(\pi r_1^2)$ per wire unit length. Hence, total self-inductance per unit length of a superconducting cylindrical wire above superconducting ground plane is

$$L_l = \frac{\mu_0}{2\pi} ln \frac{2(h_1+\lambda)}{r_1} + \frac{\mu_0}{8\pi} + \frac{\mu_0 \lambda_1^2}{\pi r_1^2} \quad (16)$$

as was demonstrated in [29].

If $r \gg \lambda_1$, magnetic flux is expelled from the wire except the sheath of thickness $\lambda_1$, and the self-inductance is

$$L_l = \frac{\mu_0}{2\pi} ln \frac{2(h_1+\lambda)}{r_1-\lambda_1}. \quad (17)$$

Consider now two parallel cylindrical wires with axis-to-axis distance in the *x*-direction, $p_x$ and axes to the ground plane distances $h_1$ and $h_2$ as shown in Fig. 3. The wire radii are $r_1$ and $r_2$, and the penetration depths are $\lambda_1$ and $\lambda_2$. Magnetic flux induced into the second inductor by current $I_1$ in the first inductor, according to (13) is

$$\Phi_{21} = \frac{\mu\mu_0 I_1}{2\pi} ln \frac{R_0}{R_1} + \frac{\mu\mu_0 I_1}{2\pi} ln \frac{R_2}{R_3}, \quad (18)$$

where $R_1$ and $R_2$ are distances from the axis of the first wire and from its image to the axis of the second wire, respectively. $R_0$ and $R_3$ are distances from the axis of the first wire and from its image to the *x-z* reflection plane at $y = -\lambda$; see Fig. 3. The first term in (18) is flux per unit length induced by the current $+I_1$ and the second term is flux of the same direction induced by the image current $-I_1$, as shown in Fig. 3. Mutual inductance of the two inductors per unit length is $M_{21} = \Phi_{21}/I_1$. From Fig. 3, $R_0 = R_3$, $R_1 = [p_x^2 + (h_2 - h_1)^2]^{1/2}$, and $R_2 = [p_x^2 + (h_1 + h_2 + 2\lambda)^2]^{1/2}$. Putting these into (18) gives for per unit length mutual inductance of two circular wires above superconducting ground plane

$$M_{21} = M_l = \frac{\mu\mu_0}{4\pi} ln \left[1 + \frac{4(h_1+\lambda)\cdot(h_2+\lambda)}{p_x^2+(h_2-h_1)^2}\right]. \quad (19)$$

Repeating the same calculations for the flux $\Phi_{12}$ induced into the first inductor by current $I_2$ in the second inductor, it is easy to verify that $M_{12} = M_{21}$. We note that mutual inductance (19) does not depend on radii of the wires because its derivation assumes that magnetic flux created by wire #1 is either completely penetrates wire #2 or expelled from its bulk into the loop between wire #2 and the ground plane, preserving the total number of flux lines. In other words, it assumes that presence of wire #2 does not alter substantially magnetic field created by wire #1.

*4.1.2 Microstrips with rectangular cross section.* In superconductor integrated circuits all inductors are formed by patterning planar thin films of thickness $t$ into strips (traces) with width $w$, so that cylindrical wires are not encountered. For calculating mutual inductance of wires with arbitrary cross section, the distances $R_i$ between the centers of the wires entering expressions (18) and (19) should be replaced by the geometric mean distances $R_{gmd}$ between cross sections of the wires [34, 37]. Self-inductance can be calculated using (19) as mutual inductance of the trace with itself by replacing $p_x$ with $r_{gmd}$, the geometrical mean distance of the cross section to itself [34, 37], and putting $h_2 = h_1$.

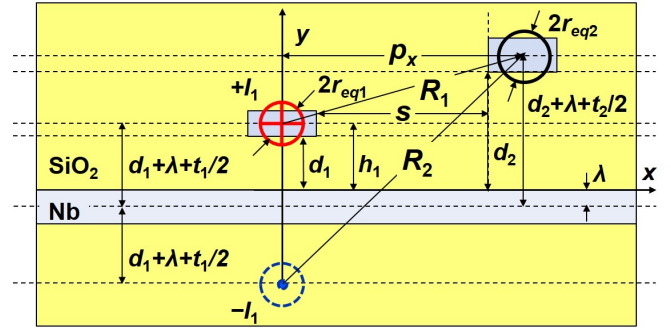

Fig. 4. Cross section of two parallel microstrips with rectangular cross section. Dielectric thickness between the microstrips and the ground plane is, respectively, $d_1$ and $d_2$. The linewidth and thickness are, respectively $w_1$, $t_1$ and $w_2$, $t_2$. For mutual and self-inductance calculations, rectangular wires can be replaced by cylindrical wires of equivalent radii $r_{eq1}$ and $r_{eq2}$ placed at the geometrical centers of the microstrips' cross sections, i.e. at $y_i = d_i + \frac{t_i}{2}$ above the ground plane surface.

For rectangular cross sections, $R_{gmd}$ can be presented as $R_{gmd} = p \cdot k(w/p, t/w)$, where $p$ is distance between their geometrical centers, and $k(w/p, t/w)$ is a weakly varying function of ratios $w/p$ and $t/w$, and mutual orientation of the cross sections, whose values are close to 1 and tabulated in [37, 66]. In all cases, $k = 1$ for $w/p \ll 1$ [37].

The typical thickness of superconductor layers is $t = 200$ nm and the smallest dielectric thickness $d_{min}$ between adjacent Nb layers is also 200 nm; see Fig. 1. Hence, the smallest value of $p$ for superconductor features located on adjacent layers is $d_{min} + t \approx 400$ nm when the second metal trace is placed right above the first one. Inductor linewidths of interest are $w \lesssim 400$ nm since we are mostly concerned with advanced deep sub-micrometer fabrication processes. The smallest pitch $p = w + s$ of inductors in the same plane depends on the fabrication process and is currently about 400 nm, whereas the minimum linewidth is about 150 nm [1]; $s$ is spacing between facing sides of the microstips. Hence, the typical ranges of inductor traces are $0.4 \lesssim t/w \lesssim 1$ and $0 \lesssim w/p \lesssim 1$.

In the given above ranges of parameters, values of $k$ in $0.92 \leq k \leq 1$ range for rectangular cross sections with long sides parallel to the line joining their geometrical centers [37]. For rectangular cross sections with long sides perpendicular to the line connecting their centers, $k \geq 1$ at all values of $w/p$ and $t/w$, increasing with $w/p$ and $w/t$. The largest $k$ value is reached at the border of the ranges $t/w = 0.4$ and $w/p = 1$ where $k = 1.0633$ [37]. Including both possible mutual orientations, the range of $k$ for inductors with dimensions typical to superconductor integrated circuits is $0.92 \leq k \leq 1.063$. Therefore, in the specified ranges of microstrip parameters, using $R_{gmd} = p$ (i.e. $k = 1$) for calculations of self-inductance does not introduce error larger than a few percent in the entire range of dimensions of interest. We will specifically discuss cases of mutual inductance of very wide and closely spaced microstrips as well as wide microstrips placed above each where accounting for $k \neq 1$ is required.

For rectangular cross section, the general expression for $r_{gmd}$ was derived by Maxwell [34] but is too long to present here. With a very high accuracy of better than 0.2%, $r_{gmd}$ can be approximated as

$$r_{rmd} = 0.2235(w + t) \quad (20)$$

in a very wide range of inverse aspect ratios $t/w$ [37]. Hence, self-inductance of superconducting microstrips with rectangular cross section, calculated as mutual inductance of the microstrip with itself, is given by

$$L_l = \frac{\mu\mu_0}{4\pi} ln\left[1 + \frac{4(d_1+t/2+\lambda)^2}{0.2235^2(w+t)^2}\right] + \frac{\mu_0 \lambda_1^2}{wt}. \quad (21)$$

In (21) we used $h_1 = d_1 + t/2$, where $d_1$ is the dielectric thickness between the wire and the ground plane, because the measured and controlled parameters in the fabrication processes are dielectric thicknesses $d_i$ between metal layers and thicknesses of metal films $t_i$; see Fig. 4. The second term in (21) is kinetic inductance of a rectangular wire per unit length, assuming uniform current distribution.

Self- and mutual inductance of conductors with any noncircular cross sections can be also evaluated by replacing them with equivalent cylindrical conductors defined as conductors producing equal far field. Any shape can be related to the equivalent circle or several circles by various transformations. Equivalent radii $r_{eq}$ for many practical as well as more academic cross sections were given in [62]. This method works very well at sufficient distances between the conductors.

For rectangular cross sections with arbitrary aspect ratio $t/w$, which is the most important case for integrated circuits and purposes of this work, a relation between $r_{eq}/w$ and $t/w$ ratio was developed in [63] and [64]. In the thin strip limit ($t \ll w$) and for the square cross section, the results are exact [64], [62]:

$$r_{eq} = w/4 \quad \text{for } t = 0 \quad (22a)$$

$$r_{eq} = \frac{aK(\sqrt{1/2})}{\pi} = 0.59017a \quad \text{for } t = w = a, \quad (22b)$$

where $K(k)$ is the complete elliptic integral. In the full range of aspect ratios $0 \leq t/w \leq 1$, results of [63] can be very accurately described by

$$\frac{r_{eq}}{w} = \frac{1}{4}\left(1 + \frac{t}{\pi w}\left(1 + ln\frac{4\pi w}{t}\right) + 0.22\frac{t}{\pi w}\left(1 + ln\frac{16\pi t}{5w}\right)\right),$$
(23a)

where the second term was given in [65] to extend the thin strip approximation [64] to $t/w \sim 0.1$ [66]. The third term is introduced here in order to describe the entire range $0 \leq t/w \leq 1$. A polynomial fitting Flammer's data [63] for $t/w \leq 1$ with less than 0.35% error was given in [40]:

$$\frac{2r_{eq}}{w} = 0.5008 + 1.0235\frac{t}{w} - 1.0230(\frac{t}{w})^2 + 1.1564(\frac{t}{w})^3 - 0.4749(\frac{t}{w})^4. \quad (23b)$$

From (16), self-inductance of microstrips with rectangular cross section is given by

$$L_l = \frac{\mu_0}{2\pi} ln\frac{2(d_1+t/2+\lambda)}{r_{eq}} + \frac{\mu_0}{8\pi} + \frac{\mu_0 \lambda_1^2}{wt} \quad (24)$$

with equivalent radius $r_{eq}$ given by (23).

In [29], we used expression (16) for self-inductance of superconducting microstrips of square cross sections. We used an effective radius $r_{sq} = a/\pi^{1/2} = 0.5642a$ giving the same cross section area, but which is 5.5% smaller than $r_{eq}$ in (22b). We found excellent agreement with experimental data and numerical simulations. We will compare more exact expression (24) with numerical and experimental results for microstrips with rectangular cross sections in the next section.

Mutual inductance of two rectangular microstrips (per unit length) with dielectric thickness, trace thickness, and trace width, respectively, $d_1, t, w_1$, and $d_2, t_2, w_2$ can be found similarly to (19) by replacing distance between centers of the wires by geometric mean distance between their cross sections using $R_{1gmd} = k_1[p_x^2 + (h_2 - h_1)^2]^{1/2}$ and $R_{2gmd} = k_2[p_x^2 + (h_1 + h_2 + 2\lambda)^2]^{1/2}$, see Fig. 4. This results in

$$M_l = \frac{\mu\mu_0}{4\pi} ln\left[1 + \frac{4(d_1+\lambda+t_1/2)\cdot(d_2+\lambda+t_2/2)}{(s+w_1/2+w_2/2)^2+(d_2+t_2/2-d_1-t_1/2)^2}\right] + \frac{\mu\mu_0}{2\pi} ln\frac{k_2}{k_1}, \quad (25)$$

where $s$ is the spacing between facing edges of the microstrips in the horizontal plane. Note that for microstrips with $t/w \leq 1$ shown in Fig. 4, $R_2 > R_1$ and, hence, $k_2 \geq k_1$. Therefore, $ln(k_2/k_1)$ in (25) is always positive, and mutual inductance of rectangular microstrips is larger than mutual inductance between their central filaments given by the first term in (25). The opposite is true in the opposite case $t/w > 1$ where the $ln(k_2/k_1)$ in (25) is negative. At $p/w \gg 1$, both $k_1$ and $k_2$ approach unity and $ln(k_2/k_1) \to 0$. For two identical microstrips ($d_2 = d_1, w_2 = w_1 = w, t_2 = t_1 = t$) spaced by $s$, eq. (25) reduces to

$$M_l = \frac{\mu\mu_0}{4\pi} ln\left[1 + \frac{4\left(d_1+\lambda+\frac{t}{2}\right)^2}{(s+w)^2}\right] + \frac{\mu\mu_0}{2\pi}(ln\, k_2 - ln\, k_1).$$
(26)

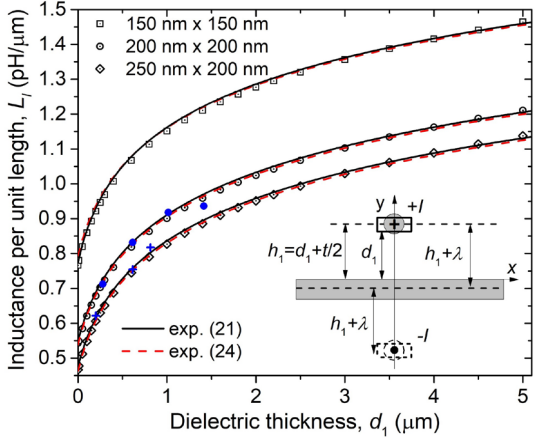

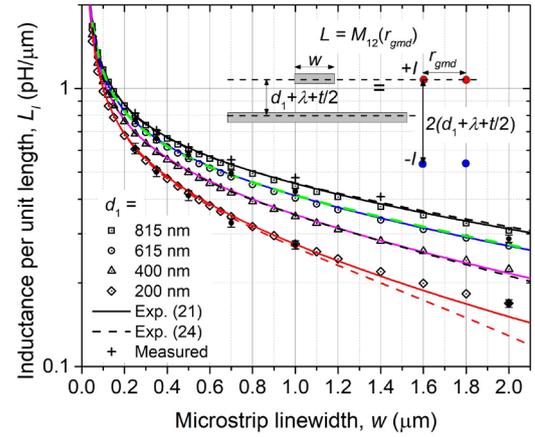

Fig. 5. Inductance per unit length of superconducting microstrips as a function of dielectric thickness. Analytical expressions (21) and (24) are shown, respectively, by solid and dash curves; numerical simulations using wxLC software and penetration depth $\lambda = 90$ nm are shown by open symbols. Microstrip dimensions are, from top to bottom, $w=t=150$ nm, $w=t=200$ nm, and $w=250$ nm, $t=200$ nm. Dash red curves – expression (24) with equivalent radius given by expression (23). Experimental data shown are: (●) – averaged data from [29], (+) – data from a recent SFQ5ee process fabrication run SFQ5A2. Inset shows schematically a rectangular microstrip with current $+I$ and its image $-I$ under the reflection plane shown by a dash line at $y = -\lambda$ below the top surface of the ground plane which is separated from the microstrip by a dielectric of thickness $d_1$. Microstrip inductance is calculated as mutual inductance of the microstrip with itself, using the geometric mean distance method, which results in (21). Microstrip inductance can also be calculated as self-inductance of a cylindrical wire of equivalent radius $r_{eq}$ producing the same far field, resulting in (24); see text.

Fig. 6. Inductance of superconducting microstrips per unit length as a function of microstrip trace width at the typical dielectric thicknesses, $d_1$ which can be realized in the MIT LL fabrication processes, from top to bottom, 815 nm (as in M7aM4 microstrips), 615 nm (as in M6aM4 microstrips), 400 nm, and 200 nm (as in M6bM7, M1aM0, etc. microstrips). Analytical expressions (21) for the GMD method and (24) with equivalent radius (23) are shown, respectively, by solid and dash curves; numerical simulations using wxLC software and penetration depth $\lambda = 90$ nm are shown by open symbols. Experimental data (+) and (●) correspond to the mean inductance values measured using circuits fabricated in all SFQ5ee process fabrication runs since 2018 to present. Standard deviation $1\sigma$ shown by error bars is less than 3% of the mean values and includes run-to-run variation and error of the measurements. Inset shows schematically a rectangular microstrip and reflection plane at $y = -\lambda$ below the top surface of the ground plane. Microstrip inductance is calculated as mutual inductance of two loops formed by cylindrical filaments spaced at $r_{gmd}$ and their images (return currents) located at distances $2(d_1+\lambda+t/2)$, as shown in the Inset, where $t$ is the microstrip thickness and $r_{gmd}$ is the geometric mean distance of its cross section to itself, which results in expression (21).

Values of $\ln k$ for various $t/w$ and $w/p$ ratios are tabulated in [37, 67]. Many useful formulas for inductance of normal metal microstrips can be also found in [68]. We will compare all the presented expressions with the results of our measurements and numerical simulations in the following section.

*4.2 Comparison with experimental data and numerical simulations*

*4.2.1 Self-inductance.* Self-inductance expression (24) with $r_{eq}$ given by (23) is plotted in Fig. 5 (by red dash curves) as a function of dielectric thickness $d_1$ for microstrips with square cross section $w = t = 150$ nm having $r_{eq} = 88.5$ nm and $w = t = 200$ nm having $r_{eq} = 118$ nm according to (22b), and of rectangular cross section $w = 250$ nm, $t = 200$ nm having $r_{eq} = 133$ nm from (23). For the same microstrip parameters, dependences (21) following from the GMD method are shown in Fig. 5 by solid black curves. Also shown are self-inductance simulated using wxLC software and experimental data obtained in this work and in [29]. In all cases considered hereafter, magnetic field penetration depth in all Nb layers (ground planes and wires) is taken as $\lambda = \lambda_1 = \lambda_2 = 90$ nm to be consistent with our prior work [27, 29, 69]. Deviations from this value will be always specified.

We can see that analytical dependence (21) based on the Maxwell's GMD method and dependence (24) following from the equivalent radius approximation are virtually indistinguishable from each other and indistinguishable from the numerical simulations and the experimental data. In [29] we showed that numerical simulations using wxLC are in excellent agreement with experimental data on per unit length inductance for all studied types of microstrip and stripline inductors. To characterize agreement between the analytical (or experimental) and the simulated results, we use the mean absolute deviation (MAD) metric

$$\chi = 100\% \cdot n^{-1}(\sum_i |(u_i - v_i)/v_i|, \quad (27)$$

where $u_i$ and $v_i$ are the analytical and numerically simulated values, respectively, and $n$ is number of data points.

In the range of thicknesses $0 \leq d_1 \leq 5$ μm presented in Fig. 5, which covers twice the range that could ever be encountered in integrated circuits and superconductor multi-chip modules ($d_1 \leq 2.6$ μm in MIT LL fabrication processes), MAD between the analytics and the wxLC numerics is $\chi=0.5\%$ for (24) and 1.0% for (21). That is, rectangular wires with small dimensions behave as cylindrical filaments for which the presented analytical expressions are exact.

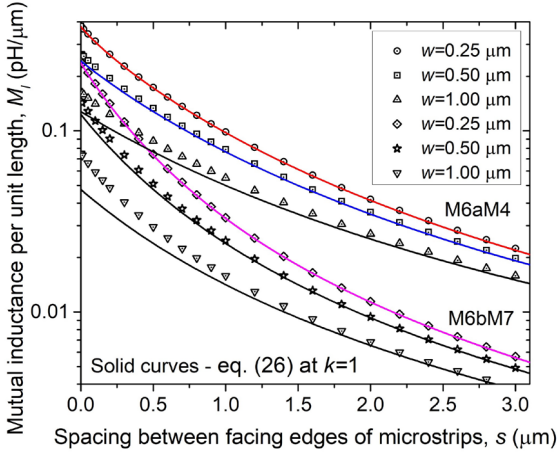

Fig. 7. Mutual inductance of superconducting microstrips per unit length as a function of spacing between them for two types of microstrips in the MIT LL fabrication processes: M6aM4 with $d_1 = 615$ nm (three top curves) and M6bM7 and equivalent types like M1aM0 with $d_1 = 200$ nm (three bottom curves). Analytical expression (26) with $k_2 = k_1 = 1$ is shown by solid curves and numerical simulations using wxLC software and penetration depth $\lambda = 90$ nm are shown by open symbols.

Next we compare dependence of self-inductance on microstrip width at a fixed Nb film thickness $t = 200$ nm for a few typical values of the dielectric thickness used in MIT LL fabrication processes, which would cover most, if not all, possible microstrips in MIT LL fabrication processes. They are: $d_1 = 200$ nm which corresponds to microstrips of $M_i a M_{i-1}$ type like M1aM0, M2aM1, etc. and inverted microstrips of $M_i b M_{i+1}$ type like M6bM7, etc.; $d_1 = 400$ nm which could be created between layers $M_i$ and $M_{i+2}$ by etching away one dielectric layer; $d_1 = 615$ nm which corresponds to the most frequently used M6aM4 microstrips in the SFQ5ee process and very close to $M_{i+2} a M_i$ microstrips; and finally $d_1 = 815$ nm which corresponds to microstrips M7aM4 and their inverted versions. All obtained results are shown in Fig. 6.

Similar to Fig. 5, we see that numerical simulations and theoretical dependences (21) and (24) are virtually indistinguishable from each other and from the experimental data in the entire range of linewidths from 50 nm to 2 μm relevant to superconductor integrated circuits, except for microstrips with $d_1 = 200$ nm. MAD metrics are: χ≤1.7% for (21) and χ≤1.9% for (24) with $d_1 \geq 400$ nm. MAD increases for $d_1 = 200$ nm to, respectively, χ≤2.6% and χ≤3.9%. In the latter case, expression (24) starts to give noticeably smaller $L_l$ values than the numerical simulations and the experimental data at $w \approx 1$ μm, whereas expression (21) starts to deviate at $w \approx 1.4$ μm. This is fully consistent with the models used. Indeed, at $d_1 = 200$ nm, distance $p$ between geometrical centers of the microstrip and its image is 780 nm, so condition $w/p \lesssim 1$ under which (21) was derived is no longer fulfilled for $w > 0.8$ μm. Expression (21) predicts lower inductance at $w/p > 1$ because it neglects function $k\left(\frac{w}{p}, \frac{t}{w}\right)$ and uses simply $k\left(\frac{w}{p}, \frac{t}{w}\right) = 1$. Accounting for $k\left(\frac{w}{p}, \frac{t}{w}\right)$ improves agreement with the numerical simulations and the experimental data at large linewidths. We will discuss this issue in relation to mutual inductance of microstrips. Similarly, expression (24) uses equivalent radius (23) describing far field. At $w > 2(d_1 + \lambda + t/2)$ this approximation becomes inadequate and predicts lower values of self-inductance.

Comparison with numerical simulations and experimental data in Fig. 6 shows that the simplest analytical expression for inductance (21) can be used for microstrips with $w \lesssim 4(d_1 + \lambda + t/2)$ and expression (24) for microstrips with $w \lesssim 2(d_1 + \lambda + t/2)$, giving accuracy of better than 2%. This accuracy is better than $1\sigma$ standard deviation of run-to-run inductance values in the existing fabrication processes for superconductor electronics. Therefore, expressions (21) and (24) provide very accurate description of self-inductance of microstrips in the range of parameters encountered in superconductor integrated circuits. They are much simpler for calculations and more accurate in this range of layer thicknesses and microstrip widths than much more complicated expressions obtained by Chang [48, 49, 70, 71]; see a comparison in [29].

*4.2.2 Mutual inductance of identical microstrips.* Dependence of mutual inductance of two identical microstrips on spacing between their facing edges is shown in Fig. 7 for a few linewidths. We selected two types of microstrips: M6aM4 with $d_1 = 615$ nm in the SFQ5ee process because of its convenience for interconnecting Josephson junctions, and M6bM7 (or any of it equivalents like M1aM0, M2aM0, etc.) with $d_1 = 200$ nm as having the smallest possible dielectric thickness and, hence, the smallest mutual coupling. Agreement between (26) at $k_1 = k_2 = 1$, which approximates mutual inductance of the microstrips by mutual inductance of their central filaments, and numerical simulations using wxLC is excellent at small linewidths. For M6aM4 microstrips, MAD is $\chi \leq 1.1\%$ for $w \leq 0.25$ μm, increasing to 3.7% at $w = 0.5$ μm, and to 10% at $w = 1.0$ μm. However, it is evident from Fig. 7 that expression (26) at $k_1 = k_2 = 1$ underestimates mutual inductance of wide microstrips at small spacings $s < w$ where $R_{gmd}/(s + w)$ becomes noticeably less than 1. The largest difference is observed at $s \to 0$. Because the largest mutual inductance is also achieved at $s \to 0$, we examine this case in more detail because of its importance for superconducting flux transformers.

Numerically simulated mutual inductance of two identical microstrips at zero spacing but without electrical contact, $s \to 0$ as a function of their width is shown in Fig. 8 along with dependence (26) at $k_1 = k_2 = 1$. Clearly, this simple approximation underestimates $M_l$ at $w/t \gg 1$ and overestimates $M_l$ at $w/t < 1$. This is consistent with the results of Maxwell [34] and Grover [37, 66] that $R_{gmd} < p$, i.e. $\ln k < 0$, at any spacing between rectangular cross sections with $w/t > 1$ and $R_{gmd} > p$ ($\ln k > 0$) at any spacing between rectangles with $w/t < 1$ when line joining geometrical centers of the cross sections is perpendicular to their longer side.

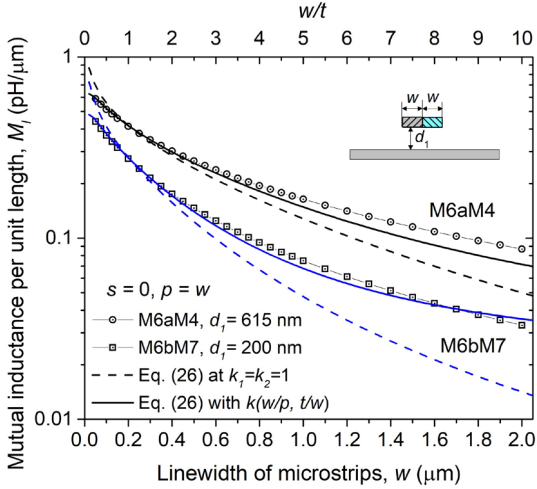

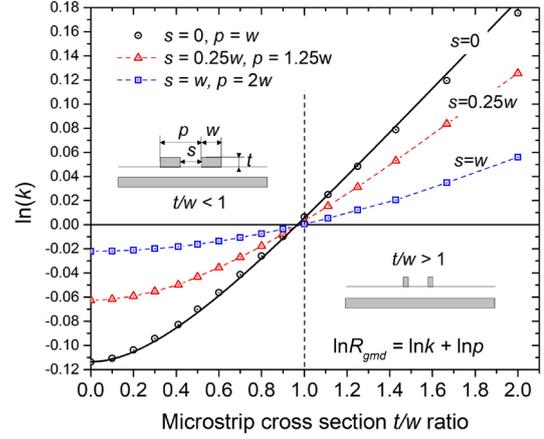

Fig. 8.  Mutual inductance of superconducting microstrips per unit length at zero spacing between them (but without electrical contact) for two types of microstrips in the MIT LL fabrication processes: M6aM4 with $d_1$ = 615 nm (three top curves) and M6bM7 and equivalent types like M1aM0 with $d_1$ = 200 nm (bottom curves). Expression (26) with $k_2 = k_1 = 1$ is shown by dash curves and numerical simulations using wxLC software and penetration depth $\lambda$ = 90 nm are shown by open symbols. Solid curves show expression (26) with $\ln(k_2/k_1) \approx -\ln k_1$ at $s = 0$ approximated by (28); see text.

Fig. 9.  Logarithm of function $k$ relating geometric mean distance of two rectangular cross sections with distance between their geometrical centers $R_{gmd} = k \cdot p$ as a function of inverse aspect ratio $t/w$ and for a few spacings between the cross sections, based on data presented in [37, 66]. Extreme case of two microstrips touching each other ($s \to 0$) corresponds to the largest values of $|\ln k|$. The latter quickly diminishes, $|\ln k| \to 0$, with increasing spacing between microstrips regardless of their aspect ratio. Also, $\ln k < 0$ at $t/w < 1$ and $\ln k > 0$ at $t/w > 1$. Solid black curve shows fitting function (28): $\ln k = -0.1135 + 0.248 (\ln(1 + t/w))^2$; see text.

Dependence of $\ln k$ on $w/t$ is shown in Fig. 9 for a few $p/w$ ratios, based on the tabulated data in [37, 66]. Independently of the aspect ratio, $|\ln k|$ reaches its maximum values in the extreme case of microstrips touching each other, $s \to 0$, and quickly decreases to zero with increasing $p/w$, i.e. with increasing spacing between rectangular microstrips. The data for $s = 0$ can be approximated with high accuracy by

$$\ln k = -0.1135 + 0.248 (\ln(1 + t/w))^2 \quad (28)$$

as shown by black solid curve in Fig. 9.

Since $R_2$ in Fig. 3 and Fig. 4 is always larger than $R_1$ and significantly larger than $w$ in all practical cases, we can neglect $\ln k_2$ in (26) because $k_2 \approx 1$ and account only for $\ln k_1$. This is done in Fig. 8 where eq. (26) with $\ln k_2 = 0$ and $\ln k_1$ given by (28) is plotted by solid curves. Account for $-(\mu\mu_0/2\pi) \ln k_1$ term in (26), which contributes $0.2|\ln k_1|$ in pH/µm to mutual inductance, significantly improves agreement with the numerically simulated values for all linewidths. The results are indistinguishable from each other at $w \lesssim 0.6$ µm.

A small difference between the numerical simulations and analytical results, clearly visible in Fig. 8 for M6aM4 microstrips at $w \gtrsim 1$ µm, remains because the analytical approach used does not take into account Meissner effect in the microstrip traces. Indeed, we explicitly took into account only superconductivity of the ground plane when we calculated positions of the trace images, whereas all other derivations for $L_l$ and $M_l$ were the same as they would be for normal metal traces with uniform current distribution. Superconductivity of traces was only accounted for by adding kinetic inductance term in (16) and (21) for self-inductance, which is irrelevant for mutual inductance. As we see, this approximation works really very well, agreeing with experimental data and numerical simulations within a couple of percent, for all microstrips with $w \lesssim 1$ µm, the range of interest for superconductor integrated circuits. However, it is not sufficient for calculating mutual inductance of wider traces with $w/t \gtrsim 5$.

Why mutual inductance of wide superconducting traces is somewhat larger than of normal metal traces with the same dimensions is easy to explain. Proximity to superconducting wire #2 alters symmetrical current distribution in wire #1 and symmetrical distribution of magnetic field created by wire #1 due to Meissner effect in wire #2, and vice versa. Both wires become strongly interacting. Magnetic field created by wire #1 induces Meissner screening currents in wire #2. On the side surface of wire #2 facing wire #1 and on the top surface of wire #2, these screening currents flow in the same direction as the excitation current in wire #1, while they flow in the opposite direction on the bottom surface, facing the ground plane, of the wire #2 and on its side facing empty space on the right. As a result of magnetic interaction, currents on the facing sides of the wires attract each other and, otherwise symmetrical, distribution of current along the width in wire #1 distorts and shifts towards wire #2 while magnetic field strength between the wires and under wire #1 reduces. This reduction in magnetic field results in reduction of self-inductance of wire #1 caused by proximity to wire #2. Expulsion of magnetic flux from the volume of wire #2 increases magnetic field under wire #2, between its bottom surface and the ground plane, and increases magnetic field near the side of wire #2 facing empty space. As a result, total magnetic flux threading the loop between the bottom surface of wire #2 and the ground plane increases with respect to the flux which would be induced without existence of Meissner screening currents in wire #2. Therefore, mutual inductance of wide superconducting wires is somewhat larger than mutual inductance given by expressions (25) and (26) which do not account for screening currents. Similarly, proximity to other superconducting inductors (features) reduces self-inductance with respect to isolated inductors, whereas expressions (21) and (24) do not account for this proximity effect.

*4.2.3 Mutual inductance of microstrips on different layers.* An important configuration of inductors for making superconducting transformers is microstrip inductors spaced vertically and placed above each other, because it allows to achieve large mutual coupling without the need to fabricate very closely spaced inductors on the same layer as described in 4.1.4. Instead, inductors are formed on different superconducting layers and share the same ground plane. Because of a close proximity of inductor traces in the *y*-direction, this arrangement is more difficult for analytical treatment than the previously considered side coupling.

We will consider in some detail coupling of M5aM4 microstrip inductors to M6aM4 microstrip inductors because this is the most convenient combinations for making compact transformers for SFQ and QFP logic cells due to proximity of these layers to Josephson junctions. The general expression (25) applies, and for microstrips of the same width $w$ placed symmetrically above each other $s = -w/2$. For this inductors $d_1$=200 nm, $d_2$=615 nm, and, at the smallest nominal (corresponding to the SFQ5ee process Design Rules) distance $R_1$ in Fig. 4 is $d_2 - d_1 + (t_2 - t_1)/2 = 0.448$ μm, and the smallest nominal distance $R_2$, between the center of the M5 trace image below the ground plane to the center of M6 trace, is $d_1 + d_2 + \lambda + t_2/2 = 1.163$ μm. Hence, for microstrips of width $0.2\ \mu m \lesssim w \lesssim 0.4\ \mu m$, $ln(k_2/k_1)$ term in (25) is small and can be neglected. Since the narrowest microstrips provide the largest mutual inductance and the smallest area of transformers, we will look at them in more detail.

Dependence of mutual inductance of M5aM4 and M6aM4 microstrips on horizontal distance $p_x$ between their geometrical centers is shown in Fig. 10 for $w_1 = w_2 =$ 0.25 μm, the minimum linewidth allowed in the SFQ5ee process, and for $w_1 = w_2 = 0.35$ μm. Experimental data are shown by open squares. As can be seen, there is an excellent (with $\chi = 1\%$) agreement between numerical simulations and expression (25) with $k_1 = k_2 = 1$, and with the experimental data. The mutual inductance is maximized at $p_x = 0$, but the maximum is not sharp, so small (up to ±0.1 μm) fabrication-caused misalignments between the microstrips do not significantly affect coupling between them. We note again, that superconductivity of considered wires does not affect their mutual inductance, i.e., narrow superconducting wires have the same mutual inductance as normal metal wires of the same dimensions placed at the same distances above a superconducting ground plane with penetration depth $\lambda$.

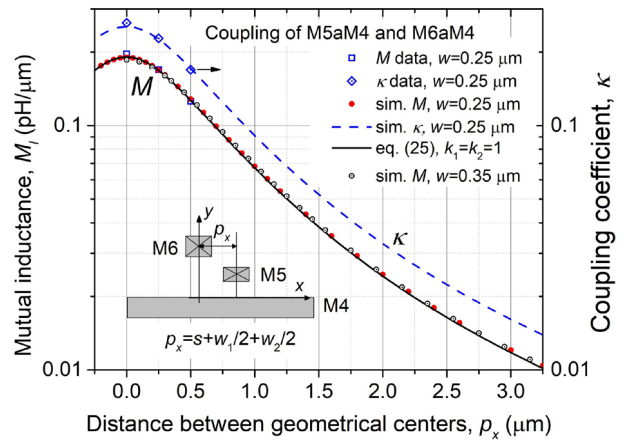

Fig. 10. Mutual inductance per unit length and coupling coefficient $\kappa = M/(L_1 L_2)^{1/2}$ between microstrips M6aM4 and M5aM4 of equal width $w$ as a function of distance between geometrical centers of their cross sections $p_x$ in the horizontal plane: (□) and (◊) experimental data for $M_l$ and $\kappa$, respectively; (●) and dash curve are numerical simulations using wxLC with $\lambda$=90 nm for $w$=0.25 μm; solid black curve – expression (25) with $ln\ k_1 = ln\ k_2 = 0$; open dots (○) are numerical simulations using wxLC for $w$=0.35 μm. Inset shows a sketch of superconducting microstrips M6 and M5 above an infinite superconducting ground plane M4. Measured self-inductances of both microstrips agree with (21) and numerical simulations within 1%. The largest deviation of the measured mutual inductance from (25) and numerical simulations is 3%.

It is clear from (19) and (25) that mutual inductance increases with decreasing dielectric thickness between inductor traces (in this case between layers M5 and M6) and reaches maximum value at $d_2 = d_1 + t_1$ when the inductor traces touch each other without making electrical contact. Hence, the main source of potential variations of mutual inductance in such vertical transformers is variation of dielectric thickness, mainly difference $d_2 - d_1$.

In integrated circuit fabrication, the main source of dielectric thickness variations is dielectric CMP which usually has a linewidth and underlying metal density-dependent planarization rate, dishing, and other topography effects. As a result, dielectric thickness above large pieces of metal in circuits (such as ground planes) and the next metal layer, e.g.,

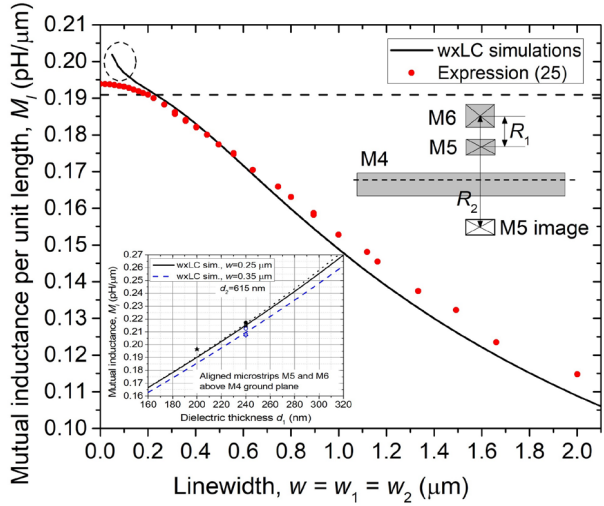

Fig. 11. Mutual inductance per unit length between aligned microstrips M5aM4 ($d_1 = 200$ nm) and M6aM4 ($d_2 = 615$ nm) of the same width as a function of their width $w$: solid line – simulations using wxLC and $\lambda=90$ nm; dash horizontal line – expression (25) with $\ln k_2 = \ln k_1 = 0$; (●) – full expression (25) using values of $\ln k_2$ and $\ln k_1$ from Table 2 in [37].

Bottom Inset: simulated using wxLC dependences of the mutual inductance on dielectric thickness $d_1$ (between M5 and M4) at a constant value of $d_2 = 615$ nm between M6 and M4 for $w = 0.25$ μm (black solid curve) and $w = 0.35$ μm (blue dash curve). Experimental data: black solid symbols for $w = 0.25$ μm; open blue symbols for $w = 0.35$ μm. Black dot curve – expression (25) at $\ln k_2 = \ln k_1 = 0$.

Top Inset: schematic cross section of the aligned microstrips, showing also image of the M5 trace below reflection plane indicated by a dash line and specifying distances $R_1$ and $R_2$ used in the text.

dielectric I4 between M4 and M5, may be somewhat different from the process target value (200 nm in this case) and differ from dielectric thickness between two narrow lines on the same metal layers. However, when dielectric layer I5 on top of M5 is processed, presence of a large ground plane on M4 layer does not affect its planarization. It is determined by other factors such as area density uniformity of Josephson junctions, in this particular example. Therefore, potential variation of multilayer dielectric thickness between distant metal layers, e.g., M4 and M6, formed by multiple depositions and multiple dielectric CMPs, are much less significant.

In order to quantitatively examine importance of thickness variation in vertical transformers, in Fig. 11 (Inset) we presented dependence of mutual inductance between vertically aligned inductors M5aM4 and M6aM4 on dielectric thickness $d_1$ while keeping $d_2 = 615$ nm constant and equal to the nominal (process target) value. Simulations were done for two linewidths $w$=0.25 (solid curve) and 0.35 μm (dash curve). Experimental data, taken at a few locations on the wafer, are shown by solid symbols for $w$=0.25 μm microstrips and by open symbols for $w$=0.35 μm microstrips. The data are in a good agreement with the simulations; small scattering of data points appears to be caused by uncertainty in the exact dielectric thickness which was measured using cross sections made by focused ion beam (FIB) and evaluated from self-inductance measurements of the microstrips. The slope $\delta M_l/\delta d_1$ at the nominal thickness $d_1 = 200$ nm for both shown linewidths is about 0.594 pH/μm² which is more than a factor of ten larger than the slope $\delta M_l/\delta w = -0.046$ pH/μm², indicating dominating effect of process-induced dielectric thickness variation over linewidth variation on parameter uniformity of vertical transformers.

Finally, we examine dependence of mutual inductance between aligned ($s = -w$) microstrips M5aM4 and M6aM4 on their linewidth. Expression (25) at $s = -w$ does not contain explicit dependence of $M_l$ on $w$, which only appears implicitly in dependences of $\ln k_2$ on ratios $w/R_2$ and $t/w$, and of $\ln k_1$ on ratios $w/R_1$ and $t/w$. Both $\ln k_2$ and $\ln k_1$ are positive and $\ln k_2 \leq \ln k_1$ for this arrangement of microstrips (shown in Fig. 11, top Inset) because line connecting centers of their cross sections is perpendicular to their long sides [37]. Taking both logarithms into account, using tabulated data in [37, Table 2] and [66, Table III], results in the dependence shown in Fig. 10 by solid dots (●). A small scattering of the points is caused by finite granularity of the data in [37] and [66]. Results of numerical simulations using wxLC are shown by a black solid curve. The largest difference between both dependences is 5% and occurs at large widths $w \gtrsim 2$ μm where $w/R_1 \gtrsim 4$.

A noticeable difference between the analytical dependence and numerical simulations appears at very small linewidths and increases with decreasing the linewidth in the region $w < t_1, \lambda$ circled by a dash in Fig. 10. We could not find any physical reason for increasing $M_l$ at $w \to 0$, and concluded that it is probably caused by some artifact of numerical simulations on a very fine grid. Analytical solution (25) is exact at $w/t \to 0$ and shows a smooth saturation of mutual inductance at a constant level in this region, as shown in Fig. 11.

We note that mutual inductance of vertically spaced microstrips in the considered type of transformers is lower than mutual inductance of their central filaments. Also, as clear from Fig. 11, superconductivity of the microstrip traces has negligible effect on their mutual inductance in the range of microstrip dimensions relevant to superconductor integrated circuits and considered here.

It is very important to note that $M_l$ dependence on horizontal and vertical separation between microstrips is logarithmical, i.e., coupling of microstrips is long-range and slowly decaying without any characteristic scale. Therefore, using microstrips in logic cells of dense integrated circuit is highly inadvisable. Using pairs of microstrips in transformers is also highly problematic in circuits requiring high area density of transformers, such as AQFP, RQL and superconducting qubit circuits, because of a significant mutual coupling between adjacent transformers. Some of these problems can be mitigated by using striplines for inductors and transformers with comprising wires placed between two parallel ground planes.

## 5. Theoretical, numerical, and experimental data for superconducting stripline inductors

*5.1 Self- and mutual inductance of superconducting striplines*

Striplines for rf and microwave transmission became widely used since 1950s due to their advantages over microstrips. Therefore, literature devoted to analysis and synthesis of striplines made of normal metals is as enormous as literature on microstrips, and we cannot reference even a tiny part of it. Many useful formulas and references are given in [38], [40], [68] whereas we only consider basic cases relevant to superconductor integrated circuits.

5.1.1 Cylindrical wire between two ground planes. Similarly to microstrip configuration, inductance of a wire between two infinite ideal (with zero magnetic field) penetration depth) ground planes can be found using the theory of images. The main difference and complication is that two ground planes, like two mirrors, create an infinite set of image currents, as shown in Fig. 12. Positive image currents $+I$ (flowing in $+z$ direction) are located at $y = \pm 2nH + h_1$ and negative (flowing in $-z$ direction) image currents $-I$ are located at $y = \pm 2nH - h_1$, where $H$ is distance between the ground planes, $h_1$ is distance from the bottom ground plane to the center of cylindrical wire of radius $r_1$ carrying positive current in $+z$-direction, $n = 0, 1, 2 ...$ is integer, and $n = 0$ corresponds to the position of the original wire. To find self-inductance of the wire and mutual inductance with another wire, magnetic fields (and fluxes) created by the original current and all image currents need to be summed up. A solution of an equivalent electrostatic problem of finding capacitance per unit length, $C_l$ of a uniformly charged cylindrical wire of radius $r$ between two grounded metal planes has been known at least since 1921 [72]. Then, inductance per unit length can be found using the standard relation

$$L_l = \frac{1}{v_{ph}^2 C_l}, \quad (29)$$

where $v_{ph} = (\varepsilon\varepsilon_0\mu\mu_0)^{-1/2}$ is the speed of light on the transmission line; $\varepsilon$ and $\varepsilon_0$ are, respectively, the relative permittivity of the dielectric between the ground planes and the absolute permittivity of vacuum; $(\varepsilon_0\mu_0)^{-\frac{1}{2}} = c$ is the speed of light in vacuum. This gives the following self-inductance

$$L_l = \frac{\mu\mu_0}{4\pi} ln \frac{cosh\frac{\pi}{H}r_1 - cos\frac{\pi}{H}2h_1}{cosh\frac{\pi}{H}r_1 - 1}. \quad (30)$$

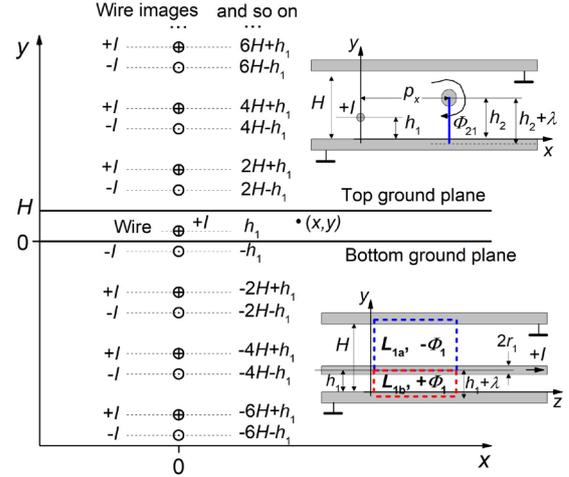

Fig. 12. Method of images for a current-carrying wire between two ideal (with zero magnetic field penetration depth) ground planes connected together at infinity. Current $+I$ in the cylindrical wire creates an infinite set of image currents caused by multiple mirror reflections off the bottom and off the top ground planes. Only a few of these image currents are shown. The $(x, z)$ plane of the coordinate system coincides with the surface of the bottom ground plane. The top ground planes is at $y = H$ and center of the wire is at $x = 0, y = h_1$. All positive image currents are located at $y = \pm 2nH + h_1$ and all negative image currents are located at $y = \pm 2nH - h_1$, where $n = 0, 1, 2 ...$ and $n = 0$ corresponds to the location of the original current-carrying wire. To find vector potential and magnetic field created by the wire current at any point with coordinates $(x, y)$ between the ground planes, vector potentials created by each current in this infinite set need to be summed up, resulting in expression (32); see text.
Top Inset shows cross section of two parallel wires of different diameter placed between two ground planes at different distances to them. $\Phi_{21}$ is the flux induced by the current in wire #1 though the loop between the wire #2 the bottom ground plane, side of which has length $h_2 + \lambda$ and shown by the solid blue line.
Bottom Inset shows two contours (loops) which can be used to calculated magnetic flux and self-inductance per unit length of the wire in superconducting state: bottom loop $L_{1a}$ runs counter clockwise along the surface of the wire in the $-z$ direction at $x = r_1, y = h_1$ and inside the ground plane at $y = -\lambda$ in the $+z$ direction. Since vector potential is independent of the $z$-coordinate, contributions to the contour integral from the two vertical sides of the loop (along the $\pm y$-directions) cancel each other. Similarly, loop $L_{1b}$ runs along the top surface of the wire at $x = r_1, y = h_1$ and inside the top ground plane at $y = H + \lambda$. Both loops are threaded by the same magnetic field lines having opposite directions below (pointing in the page) and above the wire (pointing out of the page), and inducing the same amount of magnetic flux through the loops. This picture is similar to an asymmetric coaxial line.

It is instructive to find inductance of a cylindrical wire between two ideal (with zero magnetic field penetration depth) ground planes using flux calculations because it clarifies properties of inductive loops involved and will be used for finding mutual inductance between multiple wires. In the cross section of the structure in Fig. 12, we consider two loops. The first loop, $L_{1a}$ starts on the surface of the bottom ground plane at $x = r_1; y = 0$, runs a unit length in the $+z$ direction and returns along the surface of the wire at $x = r_1; y = h_1$ a unit length in the $-z$ direction. The second loop, $L_{1b}$ is the similar loop between the wire and the top ground plane. (There is an incorrect opinion that there are individual inductances associated with each of these loops, which are connected in parallel and comprise the total inductance. This becomes immediately clear when comparing with coaxial lines.) Magnetic flux $\Phi_1$ through loop $L_{1a}$ is equal to magnetic flux through loop $L_{1b}$ and has the opposite direction because all magnetic field lines passing over the wire also have to pass below the wire − they are closed loops squeezed between the two infinite ground planes connected in the *y,z* plane at −∞ and +∞.

Inductance of any structure can be found by calculating flux induced in *any loop* through which *all* field lines within the structure must pass. In our case this can be either loop $L_{1a}$ or $L_{1b}$. Magnetic flux per unit length through any of these loops can be found using the Stokes' theorem as a contour integral of the vector potential $\mathbf{A}(x,y,z)$ over the loop in the *y,z* plane. Due to the symmetry of the problem, this integral for the loops of unit length in the *z*-direction reduces to

$$\Phi_{1a} = -\Phi_{1b} = A_z(r_1, 0) - A_z(r_1, h_1) = -[A_z(r_1, h_1) - A_z(r_1, H)], \quad (31)$$

because the only nonzero component of the vector potential $A_z(x,y)$ created by current in infinitely long wire is independent of the *z*-coordinate.

In the magnetostatic approximation, vector potential created by a circular wire with current $I$ at any point with coordinates $(x, y)$ between two infinite ground planes can be found identically to the electrostatic potential of a uniformly charged wire between two ground planes calculated by Kuntz and Bayley [72], using its full analogy with electrostatic potential; see also an identical derivation done 87 years later in [73]. This vector potential is given by

$$A_z(x,y) = -\frac{\mu\mu_0 I}{4\pi} ln\left(\frac{\cosh\frac{\pi x}{H} - \cos\frac{\pi(y+h_1)}{H}}{\cosh\frac{\pi x}{H} - \cos\frac{\pi(y-h_1)}{H}}\right). \quad (32)$$

On the surface of the wire $x = r_1, y = h_1$, the vector potential is

$$A_z(r_1, h_1) = -\frac{\mu\mu_0 I}{4\pi} ln(1 + (\frac{\sin\frac{\pi h_1}{H}}{\sinh\frac{\pi r_1}{2H}})^2), \quad (33)$$

and on the surfaces of the ideal ground planes $A_z(r_1, 0) = A_z(r_1, H) = 0$. Then, calculating magnetic flux through the loop, using (31) and (33), gives an expression for external self-inductance per unit length which is identical to (30).

For a symmetrical location of the wire between the planes ($2h_1 = H$), self-inductance (30) reduces to

$$L_l = \frac{\mu\mu_0}{2\pi} ln\left(\coth\left(\frac{\pi r_1}{2H}\right)\right), \quad (34)$$

which at $2H \gg r_1$ reduces to

$$L_l = \frac{\mu\mu_0}{2\pi} ln\frac{2H}{\pi r_1}, \quad (35)$$

The latter expression appeared in electrical engineering literature 21 years later than the work [72] and credited to Frankel [74] who in 1942, using a conformal transformation instead of the method of images, calculated wave impedance of such transmission lines.

In the case of superconducting ground planes with finite magnetic field penetration depth $\lambda$, image reflection plane is located inside each of the ground planes, a distance $\lambda$ away from the surface facing the wire. Hence, distance $H$ between the ideal ground planes in all previous formulas of this section should be replaced by $H + 2\lambda$, and distance between the wire and the bottom ground plane $h_1$ should be replaced by $h_1 + \lambda$. We also need to account for kinetic inductance of the supercurrent in the wire. Then, the total inductance of superconducting cylindrical filament with uniform current distribution placed between two superconducting ground planes becomes

$$L_l = \frac{\mu\mu_0}{4\pi} ln\left(1 + \frac{\sin^2\frac{\pi(h_1+\lambda)}{H+2\lambda}}{\sinh^2\frac{\pi r_1}{2(H+2\lambda)}}\right) + \frac{\mu_0}{8\pi} + \frac{\mu_0\lambda_1^2}{\pi r_1^2}, \quad (36)$$

where $\lambda_1$ is magnetic field penetration depth in the wire, which may differ from that in the ground planes.

In the limit $H + 2\lambda \gg \pi r_1$, expression (36) reduces to

$$L_l = \frac{\mu\mu_0}{4\pi} ln\left(1 + 4\frac{(H+2\lambda)^2}{(\pi r_1)^2}sin^2(\frac{\pi(h_1+\lambda)}{H+2\lambda})\right) + \frac{\mu_0}{8\pi} + \frac{\mu_0\lambda_1^2}{\pi r^2}. \quad (37)$$

Sinusoidal term in (36) and (37) describes dependence of the self-inductance on the distance to the bottom ground plane, which is symmetrical with respect to $h_1 = H/2$ where the maximal value of self-inductance is attained. At zero spacing between the wire and any of the ground planes, i.e., at $h_1 = 0$ and $h_1 = H$, self-inductance remains finite in the superconducting state due to magnetic field penetration into the $\lambda$-sheath of the ground planes, whereas it goes to zero for the usually considered cases of ideal ground planes.

### 5.1.2 Mutual inductance of circular wires between two ground planes.

Consider the second circular wire of radius $r_2$ running in the z-direction with axis at $(p_x, h_2)$ parallel to the previously considered first wire with radius $r_1$ and axis at $(0, h_1)$, see Fig. 12 (top Inset). Using contour integral, magnetic flux induced by the first wire into the loop of unit length along *z*-axis, between the cylindrical axis of the second wire and the bottom ground plane is $\Phi_{21} = A_z(p_x, -\lambda) - A_z(p_x, h_2)$. Obviously, the same flux passes through the unit-length loop between the second wire and the top ground plane.

From (32), vector potential created by current $I_1$ in the first wire at points $(p_x, h_2, z)$ along the axis of the second wire is given by

$$A_z(p_x, h_2) = -\frac{\mu\mu_0 I_1}{4\pi} ln\left(\frac{cosh\frac{\pi p_x}{H+2\lambda} - cos\frac{\pi(h_2+\lambda+h_1+\lambda)}{H+2\lambda}}{cosh\frac{\pi p_x}{H+2\lambda} - cos\frac{\pi(h_2-h_1)}{H+2\lambda}}\right), \quad (38)$$

and vector potential is zero inside the ground plane, $A_z(p_x, -\lambda) = 0$. Then, mutual inductance per unit length is

$$M_l = \frac{\mu\mu_0}{4\pi} ln\frac{cosh\frac{\pi p_x}{H+2\lambda} - cos\frac{\pi(h_1+h_2+2\lambda)}{H+2\lambda}}{cosh\frac{\pi p_x}{H+2\lambda} - cos\frac{\pi(h_2-h_1)}{H+2\lambda}}. \quad (39)$$

If axes of the wires are on the same (x,z) plane at $y = h_1 = h_2$, the mutual inductance reduces to

$$M_l = \frac{\mu\mu_0}{4\pi} ln(1 + \frac{sin^2\frac{\pi(h_1+\lambda)}{(H+2\lambda)}}{sinh^2\frac{\pi p_x}{2(H+2\lambda)}}). \quad (40)$$

For the symmetrical location of both wires between the ground planes, $h_1 = h_2 = H/2$, the mutual inductance becomes

$$M_l = \frac{\mu\mu_0}{2\pi} ln(coth\frac{\pi p_x}{2(H+2\lambda)}). \quad (41)$$

At large distances between the wires, $p_x > (H + 2\lambda)$, the mutual inductance decreases exponentially with increasing $p_x$, as easily seen from (40) and (41),

$$M_l = \frac{\mu\mu_0}{4\pi sinh^2\frac{\pi p_x}{2(H+2\lambda)}} \quad (42)$$

with the characteristic decay length $(H + 2\lambda)/\pi$.

To conclude, we note that results presented above are totally different from an incorrect expression

$$M_l = \frac{\mu\mu_0}{4\pi}(\frac{H}{2p_x})^2$$

given in some textbooks, e.g. in [38, Walker], for the mutual inductance of two symmetrical normal-metal striplines.

5.1.3 Mutual and self-inductance of striplines of rectangular cross section. It is important to note that Maxwell's GMD method does not work for striplines because vector potential of a filament between two ground planes is very different from vector potential of an isolated filament. To illustrate this let us consider mutual inductance between a strip of rectangular cross section, carrying current $I$ (wire #1) and a cylindrical wire (wire #2) between two ground planes. We represent a rectangular wire of width $w$ and thickness $t$ as $N + 1$ parallel cylindrical filaments of diameter $t$, carrying equal currents $I_1/(N + 1)$. Each cylindrical filament creates an infinite number of images reflected of two ground planes, as shown in Fig. 11, and has vector potential given by (38). Then, using the superposition principle, vector potential of $N + 1$ filaments at distance $p_x$ from the geometrical center of rectangular wire is given by a sum

$$A_z(p_x, h_2)$$
$$= -\frac{\mu\mu_0 I_1}{4\pi(N+1)} \sum_{n=1}^{N+1} ln\left(\frac{cosh\frac{\pi(p_x - \frac{w}{2} + (n-1)t)}{H+2\lambda} - cos\frac{\pi(h_2+h_1+2\lambda)}{H+2\lambda}}{cosh\frac{\pi(p_x - \frac{w}{2} + (n-1)t)}{H+2\lambda} - cos\frac{\pi(h_2-h_1)}{H+2\lambda}}\right),$$
(43)

where we placed the center of the first filament at the front edge of the wire facing point $p_x$. So, instead of a logarithm of geometric mean distance $N^{-1}\sum ln R_i$ for the collection of filaments, we have a geometrical mean $N^{-1}\sum ln f(R_i)$ of functions of distance $f(R_i)$ to the point. Obviously the latter cannot be reduced to a logarithm of the geometric mean distance unless $f(R_i) \equiv R_i$.

Mutual inductance between a rectangular wire and a circular filament with center at $p_x$ is found using (43) and flux calculations identical to the ones given in 5.1.2. If both wires are located on the same (x,z) plane at $y = h_1 = h_2$, the mutual inductance reduces to

$$M_l = \frac{\mu\mu_0}{4\pi} \sum_{n=1}^{N+1} ln(1 + \frac{sin^2\frac{\pi(h_1+\lambda)}{(H+2\lambda)}}{sinh^2\frac{\pi(p_x-w/2+t(n-1))}{2(H+2\lambda)}}), \quad (44)$$

Filaments of wire #1 located closer to the position of wire #2 than the geometrical center of wire #1 make much larger contribution to mutual inductance (44) than filaments located behind the geometrical center because vector potential exponentially decreases with distance. As a result, mutual inductance of two rectangular stiplines is much larger than mutual inductance between their central filaments.

To calculate self-inductance of striplines of rectangular cross section, we use an equivalent wire radius $r_{eq}$ given by (23) to replace $r_1$ in expressions (36) and (37) for stripline self-inductance. This approach gives for stripline self-inductance

$$L_l = \frac{\mu\mu_0}{4\pi} ln\left(1 + \frac{sin^2\frac{\pi(d_1+0.5t_1+\lambda)}{(H+2\lambda)}}{sinh^2\frac{\pi r_{eq}}{2(H+2\lambda)}}\right) + \frac{\mu_0}{8\pi} + \frac{\mu_0\lambda_1^2}{t_1 w}, \quad (45)$$

where we used $h_1 = d_1 + t_1/2$ to reflect that superconductor film thickness $t_1$ and dielectric thickness $d_1$ between the wire and the bottom ground plane are the fabrication process parameters.

According to (39), mutual inductance does not depend explicitly on the thickness of the ground planes. This dependence is implied in the film thickness dependence of the penetration depth

$$\lambda = \lambda_0 coth\frac{t}{\lambda_0}, \quad (46)$$

where $\lambda_0$ is the bulk value of penetration depth. All niobium layers in MIT LL current fabrication processes use 200-nm films having $\lambda = 90$ nm, except for layer M5 with thickness $t_{M5} = 135$ nm. If somebody decides to use M5 layer as a ground plane, this thin-film correction gives $\lambda_{M5} = 96$ nm.

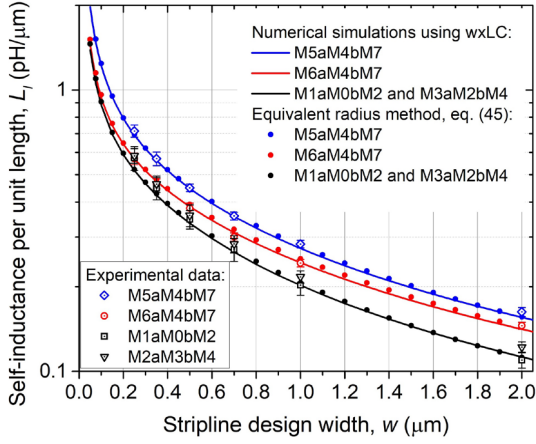

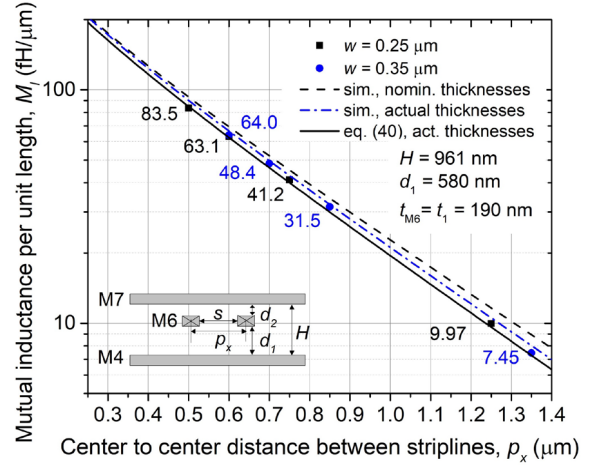

Fig. 13. Self-inductance per unit length of all main types of striplines used in the SFQ5ee process. Open symbols with error bars show mean inductance values measured on wafers fabricated in all SFQ5ee process runs from 2018 to present. Error bars represent $1\sigma$ standard deviation of each data set, including error of the measurements, cross wafer variation, and run-to-run variation of the inductance values, typically less than 3% of the mean values. Numerical simulations using wxLC are shown by color-coded solid curves while self-inductances calculated using expression (45) with equivalent radius given by expression (23) are shown by solid dots (●), (●), (●) for the same order of striplines from top to bottom: M5aM4bM7, M6aM4bM, and symmetrical striplines M1aM0bM2 and M3aM2bM4, and others geometrically identical to the latter. Nominal thicknesses of all layers in the SFQ5ee process and $\lambda$ =90 nm were used in all equations and numerical simulations: $d_1$=200 and 615 nm, and $H$=1.015 μm for M5aM4bM7 ($t_1$ = 135 nm) and M6aM4bM7 ($t_1$ = 200 nm) striplines; $d_1 = t_1$=200 nm, and $H = 600$ nm for M1aM0bM2 and M3aM2bM4 striplines. Numerical simulations agree with calculations using (45) within 1.5% in the entire range of linewidths from 50 nm to 3 μm.

Comparison with the experimental data and numerical simulations for various combinations of layer in striplines is given in the following section.

Interesting to note that mutual inductance remains nonzero even if one or both of the stiplines have zero dielectric thickness between the signal wire and any of the ground planes due to coupling via flux (currents) in the $\lambda$ sheaths.

We also note that all formulas above were derived for infinite ground planes. In real circuits, ground planes have finite width which may cause an additional coupling of microstrips and striplines by magnetic flux leakage around the ground planes. This is the so-called external mutual inductance; see e.g., [75] and references therein. In superconducting integrated circuits using several ground planes this external coupling is mitigated by connecting edges of ground planes with superconducting vias, thus forming superconducting boxes around logic cells and transmission lines. This boxing was implemented in all our circuits for mutual inductance measurements. Therefore, we will not consider external mutual inductance, which is extremely small and unmeasurable in our circuits.

*5.2 Experimental data and numerical simulations*

Fig. 14. Measured mutual inductance per unit length between stripline inductors M6aM4bM7 with the signal conductor widths $w$=0.25 μm (■) and $w$=0.35 μm (●) as a function of distance between geometrical centers of the signal conductors, $p_x = s + w$. The corresponding numerical values are shown in black for $w$=0.25 μm and blue color (for $w$=0.35 μm) for convenience. Dash and dash-dot curves show numerically simulated dependences using, respectively, the nominal thicknesses of all layers and the actual thicknesses given in Table I and obtained from SEM measurements of the inductors' cross sections. Simulations were done using wxLC and magnetic field penetration depth $\lambda = 90$ nm for all niobium layers. Solid line shows mutual inductance given by expression (40) with the actual thicknesses $d_1$ and $H$ shown in the Inset along with a schematic cross section. The characteristic decay length of mutual inductance is $(H + 2\lambda)/\pi = 0.363$ μm.

*5.2.1 Self-inductance.* Numerous experimental data for self-inductance of various striplines in the fabrication processes available at MIT LL and their comparison with numerical simulations using wxLC were given in [27] and [29]. They were found to be in a very good (with $\chi \lesssim 1$%) agreement with wxLC simulations using the actual dielectric thicknesses and the linewidths measured from secondary electron microscope (SEM) images of inductors' cross sections. In this respect, values of $L_l$ simulated using wxLC can be considered a "gold" standard. Here, we will compare expressions (43) and (44) with numerical simulations using wxLC and a subset of data obtained from multiple runs of the SFQ5ee process [30, 31] done from 2018 to present. These experimental data, averaged over all runs, are shown in Fig. 13 by open symbols with error bars for three most frequently used types of inductors: M5aM4bM7 and M6aM4bM7 striplines because they are convenient for interconnecting Josephson junctions located on layer J5 between layers M5 (junction base electrode) and M6, and M1aM0bM2 striplines and their equivalents M2aM1bM3 and M3aM2bM4, all of which are used for data and clock transmission in the bottom layers of the SFQ5ee process. Their nominal parameters are: $H = 1.015$ μm, $d_1 = 200$ nm and 615 nm for, respectively, M5aM4bM7 ($t_1 = 135$ nm) and M6aM4bM7 ($t_1 = 200$ nm); and $H = 600$ nm, $d_1 = t_1 = 200$ nm for symmetrical striplines on the bottom metal layers.

TABLE I
ACTUAL LAYER THICKNESSES OF MEASURED STRIPLINE INDUCTORS

| Layer / Inductor | M4 (nm) | M5[a] (nm) | M6 (nm) | M7 (nm) | $d_1$ (nm) | $d_2$ (nm) | H (nm) |
|---|---|---|---|---|---|---|---|
| M5aM4bM7 | 193 | 135 | n/a | 210 | 194 | 634[b] | 961 |
| M6aM4bM7 | 193 | n/a | 190 | 210 | 580 | 191 | 961 |
| M6aM5bM7 | n/a | 135 | 190 | 210 | 250 | 191 | 631 |
| Linewidth |  | 240 |  |  |  |  |  |

[a]Thickness of Nb metal remaining after anodization of the Nb/Al bilayer
[b]Total thickness from metal to metal, including anodized surface of M5

Calculations using eq. (45) with the equivalent radius given by eq. (23) are shown by small solid dots along with numerical simulations using wxLC software (solid curves). It can be seen that the results based on extremely simple expression (45) are indistinguishable from the results of lengthy numerical simulations − MAD parameter $\chi \leq 1.5\%$ for the entire range of the linewidth studied − and the experimental data. The fact that equivalent radius concept works better for wide striplines than for wide microstrips (compare Fig. 13 to Fig. 6) is not surprising because the method was specifically developed for geometries with two ground planes; see [63-65].

We also performed similar measurements for inductors M5aM4bM9, M6aM4bM9, M7aM4bM9 and M8aM4bM9 in our newer SC1 / SC2 fabrication processes having a much larger value of $H = 1.815$ μm and very different values $d_1 = 260$ nm, 655 nm, 1.055 μm, and 1.455 μm for these striplines, respectively, than in the SFQ5ee process. We have found the same excellent agreement with numerical simulations using wxLC and with expression (45) as described above, but do not present these data due to lack of space. Sensitivity of the self-inductance of striplines and microstrips to geometrical parameters $d_1$ and H diminishes with decreasing linewidth due to a rapidly growing contribution of kinetic inductance which becomes dominant at $w \lesssim 150$ nm.

*5.2.2 Mutual inductance of two M6aM4bM7 striplines in SFQ5ee process.* Due to a very large number of possible combinations, we consider here only mutual inductance of inductors of the same types as in Sec. 5.2.1 and having equal widths of the signal conductors, $w_1 = w_2 = w$. Inductors M5aM4bM7 and M6aM4bM7 are the typical case of highly asymmetric placement of the signal strip between the ground planes and convenient for testing expressions (39) and (40).

The typical mutual inductances (per unit length) $M_l$ of M6aM4bM7 inductors fabricated in the SFQ5ee process are shown in Fig. 14 for inductors with the design width $w = 0.25$ μm (■) and $0.35$ μm (●) as a function of the design distance between their geometrical centers, $p_x = w + s$. For potential users, we also give numerical values in fH μm$^{-1}$ (1000 fH = 1 pH) near each data point (in black for $w = 0.25$ μm and in blue for $w = 0.35$ μm).

Dependence $M_l(p_x)$ simulated using the above given nominal thicknesses of all layers, $\lambda = 90$ nm, and wxLC software [54-57] is shown by a dash line. It is close to the measured data but gives systematically larger values of $M_l$; the difference increases with $p_x$. Overall MAD between the measured and simulated values is $\chi$=12.1% and 16.7% for, respectively, $w = 0.25$ and 0.35 μm. At the same time, the simulated self-inductance at these linewidths $L_{sim} = 0.5677$ and 0.4719 pH/μm agrees perfectly with the measured $L_l$: $\chi$=0.4% and 1.5% for, respectively, $w = 0.25$ and 0.35 μm. Results of [29] also confirm that wxLC inductance simulator gives agreement within 1.5% with the measured self-inductances (which is the accuracy of the measurements) for all uniform transmission line-type structures studied. So, the observed difference cannot be solely caused by the simulator.

Due to a strong dependence of $M_l$ on $d_1$ and $H$ in (40), it was reasonable to assume that these parameters in the fabricated circuits differ from their nominal values in the SFQ5ee process. Results of the layer thickness measurements using SEM images of the inductors' cross sections made by focused ion beam (FIB) are given in Table I. They show that $d_1$, $H$, and Nb thickness of the M6 layer, $t_{M6}$ in the fabricated circuits are indeed smaller by about 5% than the nominal thicknesses. Using the actual thicknesses in numerical simulations (shown by a blue dash-dot line in Fig. 14) improves agreement with the measured mutual inductance by more than a factor of two: $\chi$=5.8% and 6.1% for $w$=0.25 μm and 0.35 μm, respectively. At the same time, simulated self-inductance values slightly increase, reducing agreement with the measured values to $\chi$=1.8% and 1.9%.

The best agreement of the experimental data was found with the simple analytical expression (40) for $M_l$ using the measured thicknesses; it gives MAD $\chi$<2.5% for both $w = 0.25$ μm and 0.35 μm. Hence, expression (40) can be used to calibrate numerical simulators for small linewidths of mutual inductors.

The largest coupling coefficient $\kappa = M_l/L_l$ achievable in transformers using parallel M6aM4bM7 inductors with $w = 0.25$ μm is $\kappa(s = 0) = 0.352$, at zero spacing between them. At a more practical $s = 0.25$ μm, the coupling coefficient reduces to 0.156 and becomes negligibly small at $s = 1$ μm where $\kappa = 0.018$, independently of the linewidth. Hence, for superconductor circuit design purposes, mutual inductance of M6aM4bM7 inductors in the MIT LL SFQ5ee process can be neglected if $s \geq 1$ μm because, at equal currents in the inductors, coupled magnetic flux becomes less than 2% of the self-induced flux and cannot affect operation of SFQ and other types of superconductor logic cells.

*5.2.3 Mutual inductance of two M5aM4bM7 striplines in SFQ5ee process.* Inductors M5aM4bM7 have inverted structure of M6aM4bM7 inductors. The only difference is in thickness of the M5 and M6 layers. Dependence of mutual inductance of M5aM4bM7 striplines on the distance between their geometrical centers is shown in Fig. 15 for striplines with

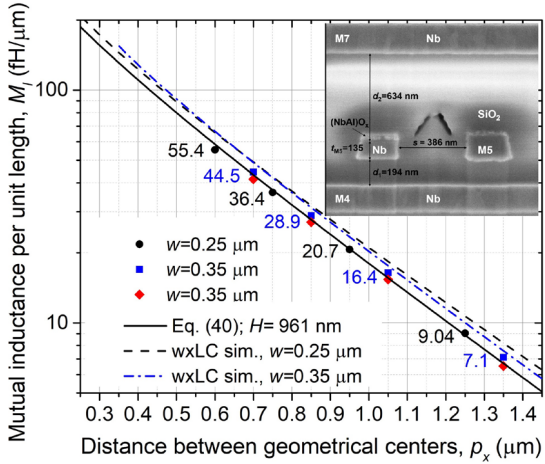
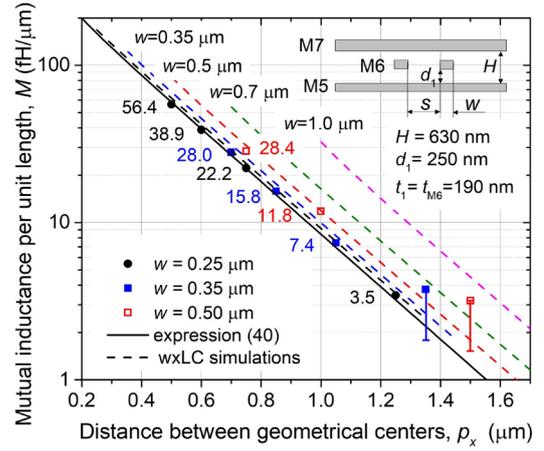

Fig. 15. Mutual inductance per unit length of two stripline inductors M5aM4bM7 with $w=0.25$ μm (●) and $w=0.35$ μm (■) and (♦), respectively, for locations C3 and E3 on the same wafer as a function of design distance between geometrical centers of M5 signal conductors. Numerical values are given for potential users. Inset: SEM image of cross section of the measured inductors with design parameters $w = 0.25$ μm and $s = 0.35$ μm; the measured thicknesses of all layers are given in Table I. Solid black line shows $M_l$ calculated using eq. (40) and the measured thicknesses $H = 961$ nm and $h_1 = d_1 + t_1/2 = 261.5$ nm. Dash and dash-dot lines are simulated dependences, using wxLC with the actual thicknesses of all layers from Table I: $t_{M4}$=193 nm, $d_1 = 194$ nm, and $d_2 = 634$ nm (instead of the nominal 680 nm); and $\lambda = 90$ nm for all niobium layers. The characteristic decay length of mutual inductance is $(H + 2\lambda)/\pi = 0.363$ μm.

Fig. 16. Mutual inductance per unit length of two stripline inductors M6aM5bM7 as a function of distance between geometrical centers of the signal conductors: (●) $w = 0.25$ μm and numerical values shown in black; (■) $w = 0.35$ μm and numerical values shown in blue; (□) $w = 0.50$ μm and numerical values shown in red Two data points at the largest distances are the upper limit on $M_l$ following from the maximum modulation current we could apply to the measuring SQUID, while the actual values are certainly lower, within the ranges indicated by "error" bars; see text. The solid line shows expression (40) using the actual thicknesses of all layers from Table II and $\lambda = 90$ nm for niobium layers M6 and M7, and $\lambda_{M5} = 96$ nm for the M5 layer. The characteristic decay length of mutual inductance is $(H + 2\lambda)/\pi = 0.26$ μm. Dash lines show $M_l$ simulated using wxLC for various linewidths, from top to bottom (from right to left): 1.0 μm (magenta), 0.70 μm (green), 0.50 μm (red), 0.35 μm (blue), and 0.25 μm (black).

$w = 0.25$ μm (●) and 0.35 μm. Two sets of data points for $w = 0.35$ μm correspond to two chips on the same wafer, C3 (■) and E3 (♦), located 44 mm apart, and show the typical repeatability of the mutual inductance. The Inset of Fig. 15 shows a SEM image of a FIB cross section of one pair of the measured M5aM4bM7 inductors; the measured layer thicknesses are given in Table I.

Mutual inductance calculated using eq. (40) and the actual thicknesses $H = 961$ nm and $d_1 + t_1/2 = 261.5$ nm is shown in Fig. 14 by a solid line. It demonstrates excellent overall agreement, $\chi = 3.3\%$ with the measured data for both linewidths. Mutual inductance simulated using the actual layer thicknesses is shown in Fig. 14 by a dash black line for $w = 0.25$ μm and by a dash-dot blue line for $w = 0.35$ μm. Using the actual thicknesses in wxLC simulations improves agreement with the measured data to $\chi$=12.9% from $\chi$=17.6% for simulations with the nominal thicknesses (not shown). Nevertheless, the simulated $M_l$ remains noticeably larger than the measured values and the values given by eq. (40).

Inspection of the cross section in Fig. 15 reveals that the actual spacing between the signal conductors is 36 nm larger than the design value, e.g., it is 386 nm instead of the designed spacing of 350 nm. This fabrication effect is caused by photolithography and etching of the M5 layer comprising Nb layer covered by a mixed (Al-Nb) oxide formed during anodization of the base electrode of Josephson junctions. After the etch step, the inner sidewalls of the conductors (sidewalls facing each other) become more vertical than the outer sidewalls which retain the typical slope of isolated M5 lines. The proximity of two closely space lines cuts off the inner corner of otherwise symmetrical trapezoidal shape of isolated M5 lines, increasing the effective spacing between the lines. If this spacing correction is also taken into account, agreement of the simulated and measured data improves, resulting in $\chi$=3.5% for $w = 0.25$ μm and $\chi$=3.6% for $w = 0.35$ μm inductors on the two chips. Simulator wxLC allowed us to account for the trapezoidal shape of inductors in Fig. 15. However, the difference of the results using the actual trapezoidal shape and the rectangular shape with the mean width was negligible and could not be even shown in Fig. 15.

The largest coupling coefficient achievable using parallel M5aM4bM7 inductors with $w = 0.25$ μm is $\kappa(0) = 0.303$ corresponding to $s = 0$. This $\kappa(0)$ is smaller than for M6aM4bM7 inductors because the latter have smaller kinetic inductance; see discussion in Sec. II and eq. (10). At $s = 0.25$ μm, the coupling coefficient reduces to $\kappa = 0.13$ and becomes negligibly small at $s = 1$ μm where $\kappa = 0.016$. Hence, for superconductor circuit design purposes, mutual inductance of M5aM4bM7 inductors in the MIT LL SFQ5ee process can be neglected if $s \geq 1$ μm, similar to the mutual coupling of M6aM4bM7 inductors.

*5.2.4 Mutual inductance of two M6aM5bM7 striplines in SFQ5ee process.* In order to investigate dependence of mutual inductance on dielectric thicknesses $H$, we measured also M6aM5bM7 inductors which have the same nominal $d_2 =$

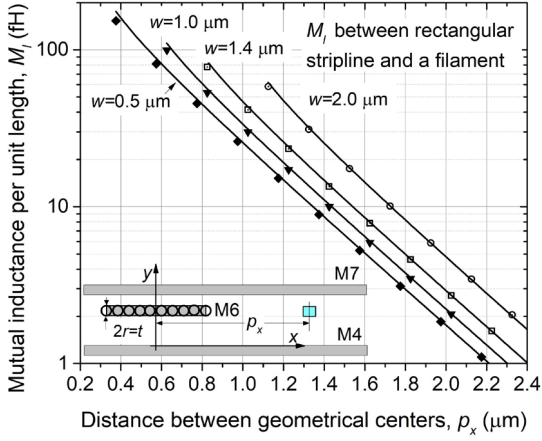

Fig. 17. Mutual inductance per unit length between a wide stripline inductor M6aM5bM7 and a filament as a function of distance $p_x$ between geometrical centers of the conductors for several linewidths, from left to right: (♦) – 0.5 μm, (▼) – 1.0 μm, (□) – 1.4 μm, and (○) – 2.0 μm. Inset shows a schematic cross section of the structure. In numerical simulations simulated using wxLC, cylindrical filament was replaced by a narrow wire with $w = 0.25$ μm on layer M6. All stripline thicknesses correspond to the SFQ5ee process: $d_1 = 615$ nm and $H = 1.015$ μm. Solid lines show expression (44) derived using mutual inductance formula (40) and replacing rectangular wire by a parallel connection of $N + 1$ cylindrical filaments, each carrying an equal fraction of the total current in the stripline. Diameter of the filaments is equal to the strip thickness $t_1 = 200$ nm, and the total number of filaments equals the integer part of $w/t_1+1$.

200 nm as M6aM4bM7 inductors in Sec. 5.2.2, but smaller $H$. The actual thicknesses measured from the cross sections were $d_1 = 250$ nm and $H = 631$ nm (see Table I).

The measured and simulated dependences of the mutual inductance for various linewidths are given in Fig. 16. Penetration depth $\lambda_{M5} = 96$ nm following from (46) was used for the M5 ground plane, replacing $h_1 + \lambda$ and $H + 2\lambda$ terms in eq. (40) with $h_1 + \lambda_{M5}$ and $H + \lambda_{M5} + \lambda_{M7}$ terms, respectively. We observed excellent agreement ($\chi < 2.5\%$) between the measured $M_l$ and calculations using expression (40), and with numerical simulations, indicating that at narrow linewidths $w = 0.25$ μm and $w = 0.35$ μm these striplines behave as cylindrical filaments. Deviations from the filament model (40) towards higher $M_l$ increase with increasing $w$ and become noticeable at $w = 0.50$ μm as shown in Fig. 16 by open squares (□) and red dash line. In all cases, mutual inductance exponentially decreases with increasing distance between the strips with a decay length of 0.26 μm.

Because $M_l$ strongly decreases with increasing spacing, modulating the measuring SQUID via the mutual inductance required applying very large currents, which started to exceed the critical current of the superconducting M6 wires used (about 25 mA) at $s \approx 1$ μm. Therefore, data points shown in Fig. 15 for 0.35-μm and 0.5-μm striplines at the largest distances are estimates of the upper limit on $M_l$, bound from above by the largest current we were able to supply to the mutual inductors, but which was not sufficient to measure one full modulation period. The actual $M_l$ values at these distances are certainly lower, laying in the range indicated by "error bars" in Fig. 16.

The measured coupling coefficients at $s = 0.25$ μm are $\kappa = 0.105$ and 0.083 for $w = 0.25$ and 0.5 μm, respectively. They reduce, respectively, to 0.042 and 0.035 at $s = 0.5$ μm, and become about 0.005 at $s = 1$ μm. Therefore, mutual coupling of M6aM5bM7 inductors in integrated circuits can be ignored for all practical purposes already at $s \gtrsim 0.7$ μm. The mutual inductance decay length is even smaller for symmetrical M1aM0bM2 and geometrically similar stripline inductors with $H = 600$ nm, about 0.25 μm. Their mutual coupling can be similarly neglected at $s \gtrsim 0.7$ μm.

*5.2.5 Mutual inductance of a wide stripline and a filament in the same plane*. Increasing $M_l$ above the filament model (40) values with increasing $w$ of rectangular strips at a fixed value of $p_x$ (see Fig. 15) was explained in 5.1.3. We consider this in more detail below. Figure 17 shows a numerically simulated dependence of mutual inductance between rectangular striplines M6aM4bM7 and a filament as a function of distance between their geometrical centers. Parameters of the striplines correspond to the SFQ5ee process: $d_1 = 615$ nm, $t_1 = 200$ nm, $H = 1.015$ μm. Instead of a circular filament we used in simulations a narrow 250 nm x 200 nm M6 wire which, as follows from the results of previous sections 5.2.1-5.2.4, is ideally described by a cylindrical filament. Results of the numerical simulations (shown by various symbols in Fig. 17) lay on parallel lines (in logarithmic scale) similar to the simulations shown in Fig. 16, indicating that increasing the linewidth at a fixed distance $p_x$ increases mutual inductance without changing its basic dependence on $p_x$. Solid lines show expression (44) following from the simplest model representing a wide strip by a parallel connection of cylindrical filaments with diameter equal to the strip thickness $t_1$, as shown in the Inset of Fig. 17. The number of filaments for each linewidth was taken as the integer part of $w/t_1+1$: three for $w = 0.5$ μm, six for $w = 1.0$ μm; eight for $w = 1.4$ μm, and 11 for $w = 2.0$ μm. As we can see, numerical results and (44) are really indistinguishable except only when the two wires touch each other. Obviously, the largest contribution to $M_l$ comes from the first filament of wire #1, closest to wire #2 due to the exponential decay of mutual inductance of filaments between two ground planes with distance between them.

By the reciprocity theorem, mutual inductance between the filament and the wire is the same if current is applied to the filament on the right of the Inset. This means that the total flux threading a wide stripline inductor can be calculated as a mean value of fluxes induced in parallel loops formed between individual filaments comprising the wide strip and the ground planes.

*5.2.6 Mutual inductance of M6aM5bM7 striplines in the SC1 process at MIT LL*. For a comparison, we studied mutual inductance of stripline inductors in our newer fabrication process, process SC1 [32], which targets minimum linewidth of 250 nm. The SC1 process differs from the SFQ5ee process by the placement of the resistor layer below the base electrode layer of JJs, layer M5. This changes thicknesses of a few dielectric layers, mainly increases the nominal thickness of layer I4 to 260 nm; see Table I. Therefore, for M6aM4bM7 inductors, the nominal parameters in the SC1 process are: $d_1 = 655$ nm, $t_1 = t_{M6} = t_{M4} = t_{M7} = 200$ nm, $H = 1.055$ μm. These changes should increase mutual inductance of M6aM4bM7 striplines with respect to the SFQ5ee process, making it easier to measure because of a larger decay length of about 0.39 μm. We also compared the obtained results with the results of mutual inductance extraction using a commercial 3D inductance extractor InductEx$^@$ 6.0 [60].

The measured mutual inductance for striplines with $w$=0.25 μm is shown in Fig. 18 along with the simulations done using wxLC (dash black line) and InductEx (colored dash-dot lines). The data were taken on two wafers, at the same location of the test chip, fabricated in the same run in order to assess wafer-to-wafer reproducibility.

It is easy to see that simple analytical expression (40), shown by a solid black line, perfectly describes the data with MAD $\chi = 1\%$, better than numerical simulations using wxLC shown by a dash black line, and significantly better than InductEx (shown by dash-dot colored lines). Simulations using wxLC give $\chi$=3.1 and 5.1% for wafers #6 and #9, respectively.

The red dash-dot line marked a) in Fig. 18 shows $M_l = M/l$ extraction from the actual test circuit layout used in the differential measurements described in Sec. II, and shown schematically in the Inset labeled a) in Fig. 17. For the extraction, InductEx requires defining ports which specify electrical connections to the inductors and ground planes. These ports were set at the ends of the inductors and marked P1, P2, P3 and P4 as shown in Fig. 18 Inset. In layout a), all ports were of M6 [M4 M7] type, supplying current to the signal conductor on layer M6 and taking it out of the ground planes M4 and M7 connected (mathematically) together at the port location, in addition to the physical connections existing around edges of the ground planes in the circuit layout. Agreement with the measurements is good: $\chi$=4.0 and 5.4% for wafers #6 and #9, respectively. Comparing with wxLC, InductEx gives about 4% larger $M_l$ at $s = 0.25$ μm and about 4% smaller $M_l$ at $s = 1$ μm; the difference is smaller in the middle. Comparing with the data, InductEx gives 7% to 10% larger $M_l$ at $s \leq 0.5$ μm and up to about 4% lower $M_l$ at $s = 1$ μm, i.e., it gives a slightly larger slope of the $M_l(s)$ dependence than in the measurements

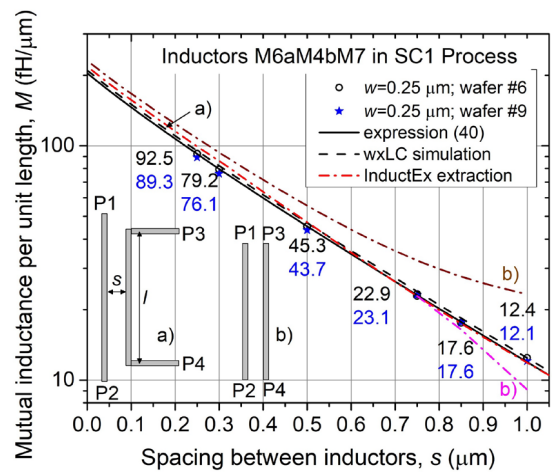

Fig. 18. Mutual inductance per unit length of two stripline inductors M6aM4bM7 fabricated in the SC1 process as a function of spacing $s$ between signal conductors: (○) and numerical values shown in black are for a test chip on wafer #6; (*) and numerical values shown in blue are for the same chip location but on wafer #9 fabricated in the same run. Inset: schematic top view of the striplines in two layouts used in the simulations. Layout a) corresponds to the measurements. Solid black line is expression (40); black dash line is dependence simulated using wxLC. All dash-dot lines are simulations using InductEx 6.0 [60] for different configurations of inductors' ports P1, P2, P3, P4, and two layout types shown in the Inset. Red dash-dot line corresponds to layout a) in the Inset and all ports of M6 [M4 M7] type; see text. Brown and magenta dash-dot curves correspond to configuration b) and different port specifications: M6 [M4] for brown and M6 [M4 M7] for magenta; see text. In layout a), ports P3 and P4 feeding current into mutual inductor #2 are located far away from stripline #1. In layout b), ports P1, P3, and P2, P4 in InductEx are placed at the ends of the corresponding striplines and spaced the same distance as the striplines; see text. Nominal thicknesses of layers in the SC1 process were used in all calculations, simulations, and numerical extractions: $d_1 = 655$ nm, $t_1 = t_2 \equiv t_{M6} = t_{M4} = t_{M7} = 200$ nm, $H = 1.055$ μm, and $\lambda$= 90 nm.

*5.2.7 Discussion of mutual inductance extraction results and caveats*. A few comments concerning mutual and self-inductance extraction are required. In wxLC, inductance matrix is calculated in the infinite line approximation of the layout type b) in Fig. 17 Inset. Electric current is applied to the signal conductor at location P1 and the return currents are extracted from the ground planes at the same location. The inductor loop is closed at port P2 located at infinity. The ground planes are mathematically connected along their edges which are parallel to the signal conductors and the signal conductors close a loop with the ground planes at infinity. This setting is unambiguous.

In InducEx, the user needs to define locations and types of ports, specifying between which layers electric current is applied. This creates some disconnect from reality because, in the test circuits, currents to the inductors under test are applied very far away from the points P1 and P3 in the layouts of types a) and b) and extracted from the ground planes also very far away from point P3 and P4, basically at the contact pads on the edges of 5-mm test chips. In the measurements, points P1 and P2 only indicate the length of the inductor being measured. Therefore, inductance and especially mutual inductance extraction results of InductEx may depend on the type of ports used. To investigate this, we used InducEx to extract mutual inductance of two long parallel striplines in configuration b) using different types of ports while keeping all other layout features, e.g., connections between the edges of the ground planes, the same as in the layout a). The uppermost dash-dot (brown) curve in Fig. 18 is the result of extraction for the layout b) when all ports are M6 [M4], meaning that the test current is applied between the signal conductor M6 and the M4 ground plane at the port location. It is clear that $M_l$ is strongly overestimated at all spacings, and at $s = 1$ μm the difference is larger than a factor of 2. The extracted self-inductance per unit length, $L_l$ is larger than the one simulated in wxLC by 8.2%.

The lowest dash-dot curve (magenta) in Fig. 18 is the extraction done with all ports defined as M6 [M4 M7], which purely mathematically connects ground planes at the locations of the four ports, in addition to the physical connections existing in the layout. This change in the ports specification, dramatically changed the extraction results: now $M_l$ agrees much better with the data for the layout b) at $s \leq 0.8$ μm, but $M_l$ becomes significantly underestimated at larger $s$. The extracted self-inductance is 4% larger than the one simulated in wxLC.

The physical and mathematical differences between the considered cases are quite significant. For the dash-dot brown curve for layout b), current excitation is applied between the M6 inductor, e.g., at port P1, and the M4 ground plane right under it. Therefore, to involve the top ground plane M7, the ground currents in the M7 plane need to flow towards its edges to vias to the M4 ground plane, from where they can flow back under the M6 inductors. This spreading out of the ground currents creates an extra inductance and significantly increases mutual coupling between adjacent M6 lines. Contrary to this, in the case of M6 [M4 M7] ports (magenta dash-dot curve), both ground planes are perfectly connected (in the software) right at the ports, along their width. These ideal connections of the ground planes near the inductors (which do not exist in the actual circuits), prevent screening currents from spreading away from the inductors and, hence, reduce mutual inductance, especially at large distances where exponentially small tails of these currents determine mutual coupling. Other possible combinations of the port types, e.g., all M6 [M7] ports or a mix of M6 [M4] and M6 [M7] ports, give mutual inductance extraction results (not shown in Fig. 18) in between the two limiting cases considered above.

In layout a) with M6 [M4 M7] ports, the problem of artificial connections between the ground planes at the ports also exists but its effect is much smaller due to a very large distance (> 20 μm) between the ports P1 and P3, and P2 and P4. Hence, distribution of the return ground currents here is determined by the layout physics and less affected by the mathematical definition of the ports. These examples demonstrate large sensitivity of the extracted mutual inductance in InductEx to seemingly small differences in specifications of the ports, which does not exist in the real circuits and real measurements.

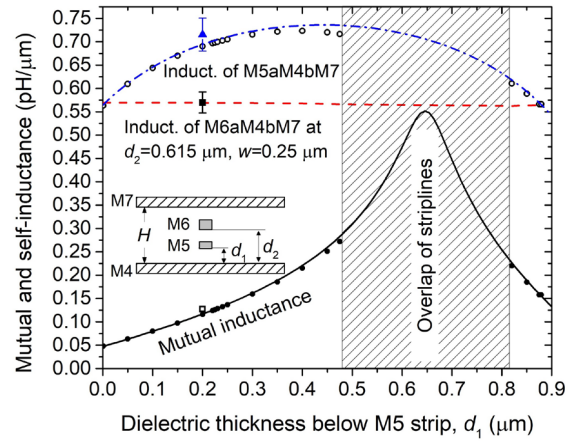

Fig. 19. Mutual inductance per unit length between vertically aligned narrow striplines M5aM4bM7 and M6aM5bM7 (see Inset) with $w = 0.25$ μm and as a function of dielectric thickness $d_1$ below the M5 strip at a fixed dielectric thickness $d_2$=615 nm below the M6 strip, corresponding to its nominal thickness in the SFQ5ee process: □ – mean value measured using test circuits fabricated in the SFQ591 fab run; solid black curve shows expression (47) for mutual inductance $M_l$; small black dots (●) are numerical simulations using wxLC and $\lambda = 90$ nm for all layers. Also shown are: measured (■) and simulated (dash red line) inductance of M6aM4bM7 striplines with M5 strip present, as shown in the Inset; measured (▲) and simulated (○) inductance of M5aM4bM7 striplines. Dash-dot blue line shows expression (45) for self-inductance of M5aM4bM7 stripline as a function of $d_1$ at $H = 1.015$ μm; equivalent radius of the M5 strip with $w = 0.25$ μm is $r_{eq} = 0.112$ μm from eq. (23). Thicknesses of all superconductors correspond to the SFQ5ee process: $t_{M5} = 135$ nm, $t_{M6} = t_{M4} = t_{M7} = 200$ nm, $H = 1.015$ μm. The largest mutual inductance is realized when surfaces of M5 and M6 strips touch each other (without making electrical contact), corresponding to $d_1 = d_2 - t_{M5} = 480$ nm and $d_1 = d_2 + t_{M6} = 815$ nm. The maximum $M_l$ lays in the physically inaccessible region of overlapping striplines (hatched region).

*5.2.8 Mutual inductance of vertically coupled M6aM5bM7 and M5aM4bM7 striplines.* Mutual inductance of two striplines is maximized when they are aligned above each other and couple vertically because exponential dependence on the distance $p_x$ is eliminated in this case. For making ac and dc flux transformers in superconductor integrated circuits, the most convenient are vertically aligned narrow striplines M6aM4bM7 and M5aM4bM7 of equal width. Their mutual inductance at $w = 0.25$ μm is shown in Fig. 19 as the function of dielectric thickness $d_1$ below M5 strip in the M5aM4bM7 striplines at a fixed dielectric thickness $d_2 = 615$ nm below M6 strips of the M6aM4bM7 striplines, corresponding to the nominal thickness in the SFQ5ee process.

For the vertically aligned strips, distance between their geometrical centers $p_x$ in expression (39) is zero and

$$M_l = \frac{\mu\mu_0}{4\pi} ln \frac{1-cos\frac{\pi(h_1+h_2+2\lambda)}{H+2\lambda}}{cosh\frac{r_{eq1}}{H+2\lambda}-cos\frac{\pi(h_2-h_1)}{H+2\lambda}}. \quad (47)$$

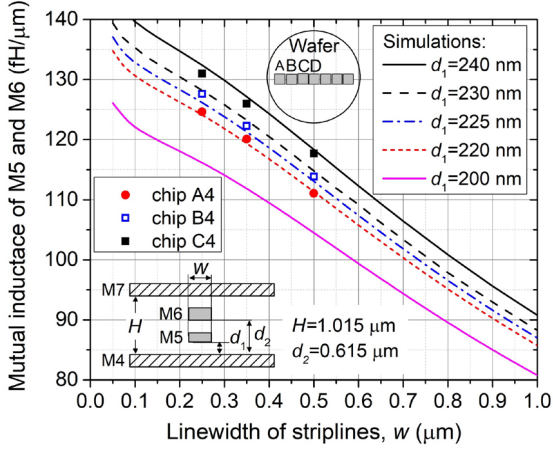

Fig. 20. Mutual inductance per unit length between vertically aligned striplines M5aM4bM7 and M6aM5bM7 (see Inset) as a function of their width $w$ for several dielectric thicknesses $d_1$ below the M5 strip at a fixed dielectric thickness $d_2$=615 nm below the M6 strip and distance between the ground planes $H = 1.015$ μm, corresponding to their nominal values in the SFQ5ee process. Experimental data shown by symbols correspond to three chips measured on the same wafer. Their locations on the wafer are shown in the top Inset. Bottom Inset shows the schematic cross section of the coupled striplines. Numerical simulation using wxLC and $\lambda = 90$ nm are shown by various lines, corresponding to different $d_1$ values. Experimental data agree better with the simulations assuming that a 10% to 20% larger $d_1$ was realized in this SFQ591 fabrication run than the nominal value of 200 nm. This is consistent with the larger measured value of self-inductance of the M5aM4bM7 striplines than the expected theoretical value, see Fig. 19.

Equivalent radii of $w = 0.25$ μm M5 and M6 strips are $r_{eq1} = 112.1$ nm and $r_{eq2} = 132.8$ nm, respectively. This dependence is shown in Fig. 19 by a solid black curve. Numerical simulations using wxLC are shown by solid dots (●) and demonstrate excellent agreement with (47). Mutual inductance reaches its maximum value in the physically inaccessible region of overlapping striplines. In any case, mutual inductance of two identical striplines is limited from above by the stripline self-inductance which is mutual inductance of the stripline with itself.

The measured self-inductances of both striplines are also shown in Fig. 19 as well as their values following from the analytical expression (45) and numerical simulations. There is a tiny difference between numerical simulations (○) and analytics (dash-dot curve) for M5aMabM7 striplines in a range of $d_1$ near $d_1 = d_2 - t_{M5_1}$, when both striplines nearly touch each other. This is because expression (45) does not take into account the presence of the M6 strip above the M5 strip. The former slightly distorts magnetic field due to the Meissner effect and slightly reduces inductance of the M5 strip. The reverse effect of the M5 strip on inductance of the M6 strip is not even visible because the former is significantly thinner than the latter.

Fig. 20 shows dependence of mutual inductance per unit length $M_l$ of vertically aligned M6 and M5 strips between M4 and M7 ground planes as a function of their width $w$. The data correspond to a few chips located along the horizontal diameter of a 200-mm wafer fabricated in the SFQ5ee process as shown in the Inset of Fig. 20. The spacing between the chips is 22 mm. Chip A is the closest to the wafer edge, and the center of chip C is about 22 mm from the wafer center. Simulated dependences on $w$ are shown by various curves corresponding to slightly different dielectric thicknesses under M5 stip. Clearly $M_l$ is much more sensitive to changes in the dielectric thickness under the stripline (see Fig. 19) than to changes in the linewidth, similarly to vertically coupled microstrips discussed in Sec. 4.2.3. This high sensitivity comes from the strong dependence of the vector potential (38) on the vertical coordinate between the ground planes, which peaks on the wire and reduces to zero inside the ground planes.

Mutual inductance increases with decreasing the linewidth and saturates at $w \to 0$ while self-inductance continues to increase as $w^{-1}$ due to increasing kinetic inductance. Hence, superconducting flux transformers should favor narrow strips despite that their traditionally defined coupling coefficient $\kappa$ linearly decreases to zero with decreasing $w$.

## 6. Effect of ground plane(s) perforations on mutual and self-inductance of microstrips and striplines

### 6.1 Self-inductance in the presence of long slits in the ground planes

Perforations in ground planes are frequently used in designing RF circuits; see, e.g. [77, 78] and references therein. In superconductor circuits they are also used to trap Abrikosov vortices created in the circuit ground planes upon cooling in a residual magnetic field [79] as well as to locally adjust (increase) inductance and mutual inductance [4, 54 , 80]. In [29], we presented data on the effect of long slits in the ground planes under microstrip and stripline inductors fabricated in the MIT LL fabrication processes We add more inductance data and a brief analysis below, and present data on mutual inductance in the presence of the slits.

If a signal trace, $L_1$ runs across the width, $g$ or along the length, $l_{slit}$ of a rectangular slit in superconducting ground plane(s), the return current, which normally flows just under the signal trace, needs to split at the point of intersection into two currents flowing around the perforation. This increases inductance of the trace and mutual inductance to a neighboring trace, $L_2$. If the trace runs along a long slit $l_{slit} \gg g$ when end effects can be neglected, we can still use per unit length self- and mutual inductances.

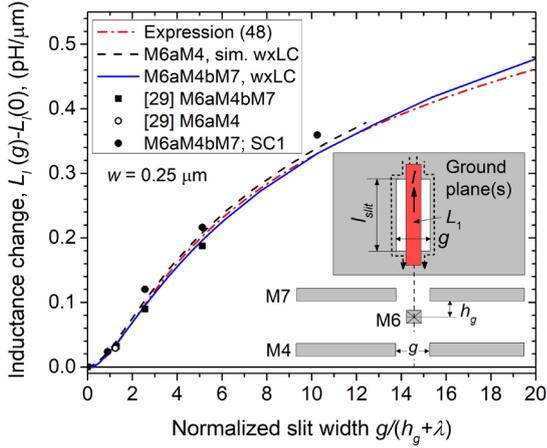

Fig. 21. Change in the self-inductance per unit length of various inductors running symmetrically along a long slit of width $g$ in the ground plane(s): (■) − $w = 0.25$ μm striplines M6aM4bM7 in the SFQ5ee process (mean values from ref. [29]); (○) – $w = 0.25$ μm microstrips M6aM4 in the SFQ5ee process (mean values from ref. [29]); (●) − $w = 0.25$ μm striplines M6aM4bM7 fabricated in the SC1 process in this work ($d_1 = 655$ nm, $H = 1.055$ μm). Dash-dot curve shows expressions (48). Solid and dash curves, respectively, for the M6aM4bM7 striplines and M6aM4 microstrips, were simulated using wxLC, nominal thicknesses in the SFQ5ee process, and $\lambda = 90$ nm. Inset: a schematic cross section and top view of an inductor with current $I$ over the slit forcing the returns currents (shown by dot cures) in the ground planes to flow around the slit. The largest slit width in experiments was 4 μm, the slit length was $l_{slit} = 30$ μm in all cases; $h_g$ is the distance from the cross section geometrical center to the nearest ground plane: 715 nm for M6aM4 microstrips and 300 nm for M6aM4bM7 striplines in the SFQ5ee process. Inductance $L_l(0) = 0.5704$ pH/μm and 0.7477 pH/μm for, respectively, M6aM4bM7 and M6aM4 inductors in the SFQ5ee process.

Inductance of the trace placed symmetrically over the slit in the ground plane (microstrip configuration) or symmetrically between congruent slits in both ground planes (stripline configuration) can be described as

$$L_l(g) = L_l(0) + \frac{\mu_0}{4\pi} \ln\left(1 + \left(\frac{g/2}{h_g + \lambda}\right)^2\right), \quad (48)$$

where $L_l(0)$ is self-inductance without the slit given, e.g., by expression (45), and $h_g$ is the vertical distance from the geometrical center of the trace cross section to the nearest ground plane. Expression (48) corrects the result given in [68] (see case J2 in Table 1 of [68]) and extends it to superconductors. The second term in expression (48) can be viewed as an aiding mutual inductance between the inductor and the slit.

Fig. 21 shows the change in the self-inductance per unit length of striplines and microstrips induced by long congruent slits with $l_{slit} = 30$ μm in the ground planes (only one ground plane in the case of microstrips) symmetrically placed along the inductors. Experimental data correspond to the SFQ5ee and the SC1 processes. Numerically simulated dependences on the normalized width of the slits are shown by solid and dash curves for the striplines and microstrips, respectively. Although the dependence on the physical width of the slit is much stronger for the striplines than for the microstrips (see ref. [29] for a comparison), both dependences collapse onto one curve after normalization of the slit width to $(h_g + \lambda)$. Dependence (48) is shown by red dash-dot curve. It describes the data very well and is virtually indistinguishable from the results of numerical simulations at $g \lesssim 6$ μm, i.e., in the entire practical range of slit widths in superconductor integrated circuits. Agreement with numerical simulations can be extended to a much wider range $g \gtrsim 10$ μm by adding a small quartic term $\beta\left(\frac{g/2}{h_g+\lambda}\right)^4$ with $\beta \approx 1/360$ under the logarithm in expression (48). Because of the weak, logarithmic dependence, at $g = 4$ μm the stripline inductance increases only by about 30%. Therefore, ground plane perforations can provide small inductance adjustments, but at the expense of dramatic reduction in density of integrated circuits.

### 6.2 Mutual inductance of horizontally spaced inductors with long slits in the ground planes

Presence of slit(s) in the ground plane(s) increases mutual inductance of all types of inductors within or near the slit. For instance, Fig. 22 shows mutual inductance of parallel M6 and M6 strips between M4 and M7 ground planes with congruent slits as a function of spacing between the signal traces. The data are for the SC1 process where $h_1 = 655$ nm and $H = 1.055$ μm for M6aM4bM7 striplines. The linewidth of inductors was $w = 0.25$ μm and the width of the slits was varied from zero to 4 μm. The mutual inductance data without the slits, $g = 0$ shown by (+) and (○) are perfectly described by the expression (40) with $p_x = s + w$ as shown by the dash line in Fig. 22 and giving $\chi = 1.5\%$.

Numerical simulations with slotted GPs are shown by dash-dot lines (using wxLC) and solid lines (using InductEx) for several values of $g$. Agreement with the wxLC simulations based on transmission line approximation is excellent: the cumulating MAD between the data for all values of $g$ and the simulated values is $\chi = 1.4\%$. However, mutual inductance extracted using InductEx6.0 from the actual layouts of the tested inductors (which include the M6 wires, the slotted GPs, and vias connecting the GPs at the periphery) is noticeably larger than the measured values in all cases. The cumulative MAD is $\chi = 12.2\%$, i.e., almost an order of magnitude larger than in the case of wxLC.

With increasing $g$, the slope of the $M_l(s)$ dependences decreases. This is equivalent to increasing the decay length $(H + 2\lambda)/\pi$ in (40), as if an effective distance between the ground planes, $H_{eff}$ increases as the slits in them widen. At $g \to \infty$, the ground planes effectively disappear $H_{eff} \to \infty$, and the exponential $M_l(s)$ dependence should turn into the standard logarithmic dependence of mutual inductance between two wires see (18). Hence, it should be possible to describe the data at $g \gg s$ by a modified expression (40) where, instead of the actual distance between the GPs, we use an effective distance $H_{eff} \sim g$ and, instead of the actual distance from the bottom GP to the trace center $h_1$, we use an effective distance $h_{1eff} \sim H_{eff}/2$

$$M_l = \frac{\mu\mu_0}{4\pi} \ln\left(1 + \frac{\sin^2 \frac{\pi h_{1eff}}{H_{eff}}}{\sinh^2 \frac{\pi(s+w)}{2H_{eff}}}\right). \quad (49)$$

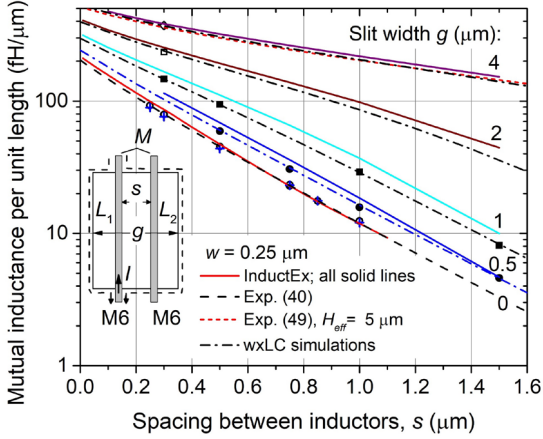
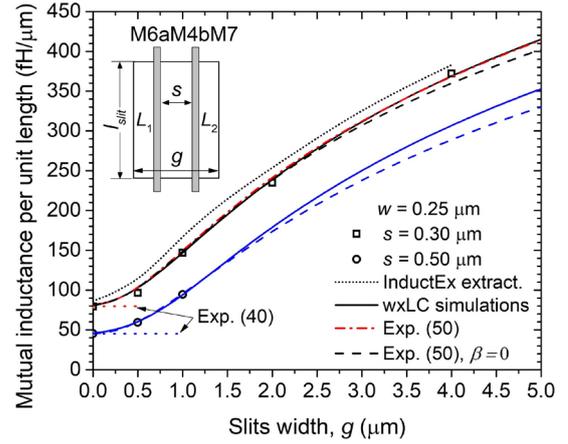

Fig. 22. Mutual inductance of parallel inductors (striplines) on layer M6, running symmetrically along the congruent slits in the ground planes M4 and M7; the slits width $g$= 0, 0.5, 1, 2 and 4 μm, from top to bottom. The slits length $l_{slit}$ = 30 μm, the inductors linewidth $w$ = 0.25 μm. The measured values are shown by various solid and open symbols, and correspond to the SC1 process (run #SFQ104191): $h_1$ =755 nm, $H$ =1.055 μm. Solid lines of various colors show the mutual inductance extracted using InductEx 6.0 from the layouts of the tested inductors, using the layer definition file corresponding to the SC1 process. Dash-dot curves show results of numerical simulations using wxLC (transmission line approximation). Black dash line for $g$ = 0 shows analytical expression (40) for the mutual inductance. Red short-dash line for $g$ = 4 μm shows expression (49) which accounts for the presence of slits by using an increased effective distances between the ground planes, $H_{eff}$ and to the stripline $h_{1eff} = H_{eff}/2$ in expression (40).

Fig. 23. Mutual inductance of parallel inductors (striplines) on layer M6, running symmetrically along the congruent slits in the ground planes M4 and M7 as a function the slits width for two values of spacing between the inductors: (□) − $s$ = 0.3 μm; (○) − $s$ = 0.5 μm. The inductors linewidth $w$ = 0.25 μm, the slits length $l_{slit}$ = 30 μm, the SC1 process (run #SFQ104191): $h_1$ =755 nm, $H$ =1.055 μm. Short dash curve shows the mutual inductance extracted using InductEx 6.0 from the layouts of the tested inductors at $s$ = 0.3 μm. Solid curves (both black and blue) show results of numerical simulations using wxLC (transmission line approximation). Dash curves show exp. (50) with no quartic term, $\beta$ = 0: $h_{1g}$ = 0.42 μm (black, for $s$ = 0.3 μm); $h_{1g}$ = 0.53 μm (blue, for $s$ = 0.5 μm). Red dash-dot line for $s$ = 0.3 μm shows Exp. (50) with $h_{1g}$ = 0.42 μm and $\beta$ = 1/219; it is indistinguishable from the results of the wxLC up to $g$ = 6 μm. InductEx overestimates mutual inductance by 10.5% on average.

Dependence (49) is shown in Fig. 22 by a short-dash red line for inductors with $g$ = 4 μm. At $H_{eff} \approx 5$ μm and $h_{1eff} = H_{eff}/2$, it is practically indistinguishable from the results of the wxLC numerical simulations.

On the other hand, at small slit widths $g \lesssim 2H$ and small spacings $s \lesssim H/2$ when increase in the mutual inductance is relatively small, mutual inductance of two parallel striplines can be very well described, similarly to the self-inductance (48), as

$$M_l(g) = M_l(0) + \frac{\mu\mu_0}{4\pi} \ln\left(1 + \frac{(g/2)^2}{h_{g1}^2} + \beta \frac{(g/2)^4}{h_{g1}^4}\right), \quad (50)$$

where $h_{g1} \sim H/2$ and $\beta \ll 1$ are fitting parameters, and $M_l(0)$ is given by (40). Dependence (50) at $\beta$ = 0 is shown in Fig. 23 by dash curves along with the experimental data (symbols), numerical simulations using wxLC (solid curves) and numerical extraction using InductEx (short dash curve). The values of $M_l(0)$ calculated using (40) are shown in Fig. 23 by dotted lines. With a small quartic term, exp. (50) is indistinguishable from the results of numerical simulations using wxLC. It is also clear that wxLC provides a much better description of the experimental data than InductEx 6.0 that overestimates mutual inductance, on average, by 10.5%.

It is also obvious from the presented data that slits, even of small width, have gigantic effect on the mutual inductance, much stronger than on the self-inductance, especially when mutual inductance is low, i.e., inductors are far apart. Hence, relatively narrow slits in the GPs can be used to make more efficient flux transformers without significant compromise of the circuit density. However, one should be very careful in designing and using wide slits (moats) for the purpose of trapping magnetic flux in them because, at the same time, they strongly enhance mutual coupling of all nearby inductors; see also [29].

### 6.3 Mutual inductance of vertically spaced inductors with long slits in the ground planes

Inductors defined on adjacent metal layers and aligned over each other are the most convenient configuration for realizing flux transformers. Their mutual inductance can also be substantially increased by slotting the GPs as shown in Fig. 24 for the case of symmetrically placed congruent slits of width $g$; see a sketch of the cross section in the Inset. We used the same long slit configuration $l_{slit} \gg g$ μm with $l_{slit} = 30$ μm and in sec. 6.2 in order to use per unit length values. The slits-induced increase in $M_l$ is the most dramatic for the striplines, more than 50% and 200 % at relatively modest slit widths, respectively, 1 μm and 2 μm. $M_l(g)$ dependences simulated using wxLC and nominal thickness of all layers corresponding to the SFQ5ee process are shown in Fig. 24 by the solid curves both for the microstip and stripline configurations.

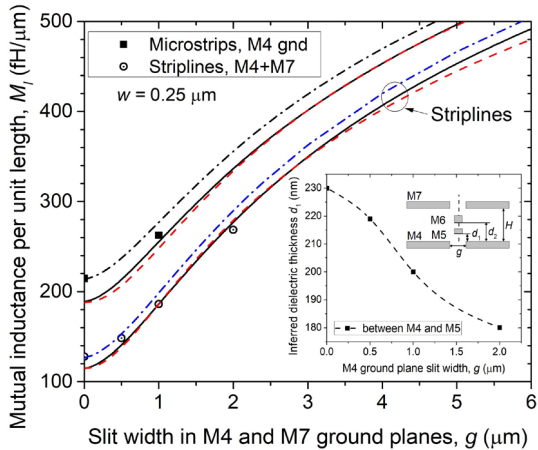

Fig. 24.  Mutual inductance of aligned inductors on adjacent metal layers M5 and M6 between slotted ground planes M4 and M7; long congruent slits are symmetrically placed along the inductors as shown in the Inset. Data (■) correspond to the microstrip configuration (single GP, layer M4) and to the stripline configurations (○) with two GPs, layers M4 and M7. The slits length $l_{slit} = 30$ μm, the inductors linewidth $w = 0.25$ μm, the SFQ5ee fabrication process run SFQ591, the same as in Fig. 20. Solid curves are numerical simulations using wxLC and the nominal parameters of the SFQ5ee process: $d_1 = 200$ nm, $d_2 = 615$ nm, $t_1 = 135$ nm, $t_2 = 200$ nm, $H = 1.015$ μm. Dot-dash curves are numerical simulations with $d_1 = 240$ nm (for microstips) and $d_1 = 230$ nm (for striplines) determined from the measured self- and mutual inductance of the inductors without slits in the GPs. Dash curves show exp. (50) at $\beta = 0$ and $h_{g1} = 460$ nm for the microstrips and at $h_{g1} = 400$ nm for the striplines. Inset: Dependence of the dielectric thickness $d_1$ between the bottom of the M5 strip in the slit and the top surface of the bottom ground plane M4 on the slit widths, inferred from the mutual and self-inductance measurements. The dielectric thickness above the slit decreases because of dishing during chemical-mechanical polishing.

It is easy to infer the actual dielectric thicknesses realized in a given fabrication run at a particular location on the wafer from the self-inductance data on the microstrips, which depend only on $h_i$, and on the striplines, which depend on both $h_i$ and $H$. This gave us an increased value of $d_1$, between 220 nm and 240 nm depending on location on the wafer, and very close to the target value of $d_2 = 615$ nm, consistent with the mutual inductance data in Fig. 20. Numerical simulations with the increased $d_1$ are shown in Fig. 24 by the dot-dash curves. They agree perfectly with the measurements at $g = 0$ but give slightly larger $M_l$ values with increasing $g$. Since there is no doubt at this point that wxLC simulations are very accurate, this discrepancy can be caused by decreasing dielectric thickness below M5 wires in the slit as a results of dishing during the CMP of the I4 dielectric above the patterned M4 layer. The inferred, from the $M_l$ values, local dielectric thickness $d_1$ is shown in the Inset of Fig. 24. Interestingly, these small changes in the local thickness and topography of the dielectric are not seeing in the self-inductance of the M6 strips because the dielectric between the M4 GP and M6 layer was deposited and planarized multiple times, whereas the I4 dielectric between M4 and M5 layers is deposited and planarized only once.

To conclude, relatively narrow slits ~ 1 μm in the ground plane(s) can increase mutual inductance of adjacent narrow wires by ≳50% both for vertically and horizontally spaced inductors. Mutual inductance of vertically spaced inductors is very sensitive to the difference $h_2 - h_1$, i.e., the difference in the local dielectric thicknesses $d_2 - d_1$, and can be used as a sensitive nondestructive method of measuring the local interlayer dielectric thickness. This high sensitivity could be a source of variability of the flux transformer using vertically coupled inductors. E.g., a 9% change in the dielectric thickness between layers M5 and M6 in the SFQ5ee process would cause a 10% change in the mutual inductance of M5 and M6 wires in the transformer. In this respect, transformers formed using closely spaced wires in the same plane could have smaller variability but would also have larger area.

## 7. Inductance of bends and meanders (serpentines)

### 7.1 Inductance of right-angled bends

Straight-line inductors described in the previous sections are widely used in logic cells and in passive transmission lines. Very often inductor bending is needed to connect JJs with different *x,y* coordinates. There is a considerable microwave engineering literature on inductance and capacitance associated with discontinuities of microstrips and striplines; see [39-40], [80-84], and references therein. Nevertheless, calculations and interpretation of the bend inductance in publications and handbooks on microwave engineering are not applicable to superconducting structures and confusing in some cases. Inductance estimates in [84] are only applicable to very wide $w \geq 2.5$ μm superconducting microstrips in the nonexistent now fabrication process.

In superconductor circuit design community, a rule of thumb has been to assign an additional inductance, $L_c$ to a corner square of the right-angled bend equal to $0.56L_{sq}$, where $L_{sq}$ is the sheet inductance (inductance per square) of the straight-line inductor. This number comes from analogy to extra resistance of the corner square of right-angled bends of thin-film resistors [85, 86]. However, we are not aware of justification for applying it to superconducting microstrip and stripline inductors. Therefore, we measured inductance of right-angled bends of stripline inductors with different widths.

Inductance of a bend is associated with redistribution of magnetic field around of and currents in the signal wire and in the ground planes, the so-called current crowding near the inner corner [85]. It results from a constructive addition of magnetic fields created by the straight runs near the inner corner.

*7.1.1 Corner inductance definition.* It would be natural to define the bend inductance as the difference between inductances of the bended and straight inductors of the same length. Instead, it has been a custom to calculate inductance associated with a right-angle bend as referenced to the corner square between lines $OQ$ and $OQ'$ in Fig. 25(a). A right-angled bend inductance, often called corner inductance, $L_c$, is defined as a difference between the total inductance of the bended inductor and inductance of the same-width straight line inductor with length $\ell_1 + \ell_2$, where $\ell_1$ and $\ell_2$ are the lengths of the horizontal and vertical sections between the inductor ends and the corner square with side $w$:

$$L_c = L_{tot} - L_l(\ell_1 + \ell_2), \quad (51)$$

where $L_l$ is the inductance per unit length of the straight line inductor.

*7.1.2 Corner inductance measurements.* Using definition (51), $L_c$ can be measured by the differential method described in Sec. 2, Fig. 2, simply placing a straight-line inductor with length $\ell = \ell_1 + \ell_2$ in the left arm of the SQUID in Fig. 1 and the bended inductor in the right arm, and determining $|L_c|$ from the SQUID modulation period. This permits measuring also mitred bends which are used in transmission lines to compensate for the effects of the discontinuity, e.g., extra capacitance; see [40], [83] and references therein. The right-angled bends measured in this work are nominally with $b = 0$, although there is always some small rounding of the inner and outer corners caused by the light diffraction in the 248-nm photolithography tool used.

To increase sensitivity of the measurements and avoid errors related to measuring small differences of two big numbers, we used meandered striplines with multiple right-angled bends spaced a distance $s$ as shown in Fig. 25(b). Parameters of the measured inductors are given in Table II.

If mutual inductance between the straight segments of inductors with bends can be ignored, self-inductances of the left and right inductors in the test structure and their difference, $L_{mes}$ which is measured using the SQUID modulation period (see Sec. 2), can be presented as

$$L_{left} = L_l(l - N_{cl}w) + L_c N_{cl}, \quad (52a)$$

$$L_{right} = L_l(l - N_{cr}w) + L_c N_{cr}, \quad (52b)$$

$$L_{meas} = |L_l(N_{cr} - N_{cl})w - L_c(N_{cr} - N_{cl})|, \quad (52c)$$

where $L_l$ and $l$ are, respectively, the linear inductance of the straight stripline and the total length of the inductors measured along the central line, i.e., the total length between the first and the last vertex if drawn as a fixed-width polyline; $N_{cl}$ and $N_{cr}$

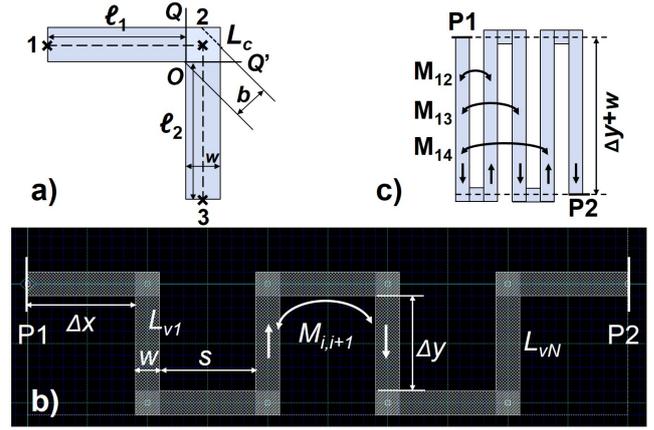

Fig. 25. (a) Right-angled bend of a signal trace of an inductor. Corner inductance is defined as referenced to a corner square between lines $OQ$ and $OQ'$. Inductance trace is drawn as a constant width $w$ polyline with vertices at points 1, 2, and 3 marked by ×. Inductance is measure between ports along sides 1 and 3. The dash line near the corner shod a metered corner, where $b$ is the corner metre, used in transmission lines to compensate for an extra capacitance associated with the corner. (b) Meandering (serpentine) inductor between ports P1 and P2, having $N_c$ corners (90-degree bends). Inductance is measured between port P1 and P2. Electric current in adjacent vertical sections of the meander flows in the opposite directions creating an opposing (negative) mutual inductance $M_{i,i+1}$ between them. Mutual inductance between the next neighbors is adding (positive). The length $\Delta x$ of the meander ends is usually equal to the spacing $s$ between the vertical sections.

TABLE II
PARAMETERS OF MEASURED SERPENTINE INDUCTORS M6AM4BM7 IN THE SFQ5EE PROCESS

| $w$ (μm) | $s$ (μm) | $N_{cr}$ [a] | $N_{cl}$ [b] | $l$, total length (μm) | $l_{vert}$ [c] (μm) | $L_{meas}$ [d] (pH) |
|---|---|---|---|---|---|---|
| 0.25 | 0.5 | 44 | 4 | 35.3 | 16.5 | 5.792 |
| 0.35 | 0.7 | 44 | 4 | 47.1 | 23.1 | 5.823 |
| 0.50 | 1.0 | 44 | 4 | 67.0 | 33.0 | 5.746 |
| 0.70 | 1.4 | 44 | 4 | 94.2 | 46.2 | 5.628 |
| 1.00 | 2.0 | 44 | 4 | 133.0 | 66.0 | 5.657 |
| 2.00 | 2.0 | 44 | 4 | 174.0 | 86.0 | 5.746 |

[a] $N_{cr}$ — number of corners in the right-arm inductor of the differential SQUID
[b] $N_{cj}$ — number of corners in the left-arm inductor of the differential SQUID
[c] $l_{vert}$ — total length of the vertical segments of the meander in each arm
[d] $L_{meas}$ — the measured difference of inductances of the right and left arms, corresponding to 40 right-angled bends (90-degree corners)

are the numbers of 90-degree bends in the left and the right inductors of the differential SQUID, respectively. This gives

$$L_c = L_{sq} - L_{meas}/(N_{cr} - N_{cl}), \quad (53)$$

where $L_{sq} \equiv L_l w$ is inductance per square.

*7.1.3 Corner inductance of striplines: results.* The $L_c$ values extracted using eq. (53) and the data in Table II for right-angled bend of M6aM4bM7 striplines of various width are shown in Fig. 26 by (+). In the calculations, we used the independently measured $L_l$ of the M6aM4bM7 striplines; see Fig. 13 and [26]. Thus extracted $L_c$ strongly depends on $w$ and becomes negative below $w = 0.25$ μm.

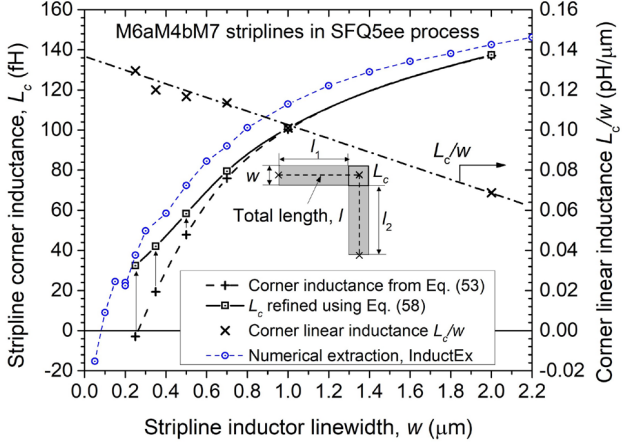

Fig. 26. Inductance of right-angled bend (corner inductance) in the signal trace of stripline inductors M6aM4bM7 with various width (□), fabricated in the SFQ5ee process, run SFQ599: $d_1 = 615$ nm, $H = 1.015$ μm. Corner inductance is referenced to the corner square of side $w$ shown in the Inset. Two serpentine inductors with the same total length $l$ but vastly different number of corners were used in the measurements. The corner inductance determined using simplified eq. (53) is shown by (+). The true corner inductance $L_c$ calculated using eq. (58) which account for the mutual inductance between parallel segments of the serpentine inductors is shown by (□). Corner inductance values extracted using InductEx from layouts schematically shown in the Inset are shown by (⊙); they are noticeably larger than the measured values.

Negative values of the corner inductance of microstrips can be found in many published theoretical and experimental results, e.g. [80, 81]. To be sure that this is not an artifact, we refined eqs. (53)-(54) to account for the mutual inductance of the nearest-neighbor vertical segments with spacing $s \leq 2$ μm. We neglected mutual coupling to more distant neighbors and the mutual inductance between the horizontal segments because their midpoints are at distances $((s + w)^2 + (\Delta y + w)^2)^{1/2} \gg 1$ μm and their mutual inductance per unit length, $M_l$ is much less than 1% of the self-inductance per unit length $L_l$; see Sec. 5.2.2. Here $\Delta y$ is the length of the straight vertical segment of the serpentine, as shown in Fig. 25(b).

Because electric current in the adjacent vertical sections flows in opposite directions, mutual inductance of the nearest neighbors is negative (opposing inductors). Then, inductance of the $i$-th vertical section of the right-arm serpentine can be presented as

$$L_{vi} = L_l(\Delta y - w) + 2L_c - M_{i-1,i} - M_{i,i+1}. \quad (54)$$

The first ($i = 1$) and the last ($i = N_c/2$) vertical segments have only one nearest neighbor. Using $M_{ij} \equiv M_{ji} = M_l \Delta y$, we sum up the inductances of the vertical sections to obtain

$$L_{vert} = L_l(\Delta y - w)\frac{N_{cr}}{2} + N_{cr}L_c - 2M_l\Delta y\left(\frac{N_{cr}}{2} - 1\right), \quad (55)$$

where we included inductance of the corners. Similarly the sum of all horizontal sections of the meaner with total length $l_{hor}$ gives simply

$$L_{hor} = L_l(l_{hor} - w\frac{N_{cr}}{2}), \quad (56)$$

because we assigned the corners to the vertical sections. The sum $L_{vert} + L_{hor}$ gives the total inductance of this type of meanders

$$L_{tot} = L_l(l - N_{cr}w) + N_{cr}L_c - 2M_l\Delta y(N_{cr}/2 - 1), \quad (57)$$

where $l = l_{vert} + l_{hor}$ is the total length of the meander along its central line. We used meanders with $\Delta y = s + w$.

Inductance of the left-arm serpentine inductor in the measurements does not require any correction with respect to the eq. (53) because it has very large values of $s \gg 5$ μm due to a small number of corners and mutual coupling between all its segments can be completely neglected. Then, the refined formula for extracting corner inductance from the measured inductance $L_{meas}$ becomes

$$L_c = L_l w - \frac{L_{meas}}{N_{cr} - N_{cl}} + \frac{2M_l\Delta y\left(\frac{N_{cr}}{2} - 1\right)}{N_{cr} - N_{cl}}. \quad (58)$$

The corner inductance obtained using this more general expression is shown in Fig. 26 by open squares (□). At small linewidths $w \leq 0.5$ μm and, accordingly, small spacings between the meander's vertical segments $s \leq 1.0$ μm, the mutual inductance correction in eq. (58) to the extracted $L_c$ values is noticeable, while it is negligibly small for all serpentine stripline inductors with $s > 1$ μm. The extracted, using eq. (58), $L_c$ values remain positive at all measured linewidths, decreasing strongly with decreasing $w$. Hence, at small linewidths of the striplines, corner inductance can be completely neglected.

The $L_c/w$ ratio, shown in Fig. 25 by (×), can be used as the linear inductance of the right-angled bend. It is a more convenient quantity for design purposes than $L_c$ because it changes much less than $L_c$, increasing only by a factor of about two while $w$ decreases by a factor of ten, from 2 μm to 0.25 μm. For designing any serpentine stripline inductor with $N_c$ corners, the target inductance $L$, and $s > (H + 2\lambda)$, the total length of the serpentine (along its central line) should be

$$l = \frac{L}{L_l} + N_c w\left(1 - \frac{L_c}{wL_l}\right), \quad (59)$$

which is, for considered striplines, always larger than the length of the straight-line inductor $L/L_l$. Hence, serpentine inductors always occupy larger area than straight-line inductors with the same linewidth.

Both $L_c$ and the ratio $L_c/(L_l w) \equiv L_c/L_{sq}$ strongly depend on the inductor linewidth; see also Fig. 26. With increasing $w$ to 2 μm, $L_c/L_{sq}$ approaches 0.5, a value close to the 0.56 value used by the superconductor electronics design community from the analogy to the corner resistance of thin film resistors [86]. However, for all the typical linewidths used in superconductor integrated circuits $w \leq 1$ μm the actual ratio which needs to be used in the circuit design is much smaller. Contribution of right-angled bends can be practically ignored at $w \leq 0.25$ μm.

*7.1.3 Numerical extraction of the corner inductance.* For a comparison, we extracted $L_c$ and $L_c/L_{sq}$ from the layouts of the striplines with bends, using InductEx 6.0. The accuracy of InductEx with the defaults settings for the SFQ5ee process was not sufficient for the extraction, giving random values of the corner inductance which changed between small positive and small negative numbers. We increased the number of filament layers (in the *z*-direction, perpendicular to the ground planes) from 2 to 3 for all Nb layers in the stripline and decreased the maximum allowed segment size (in the film plane) from 1 μm (recommended setting) to 0.15 μm. However, with these settings, we were not able to extract inductances from the actual layout, occupying 80 μm × 80 μm, of the two serpentine inductors used in the differential measurements because of insufficient computer memory and timeout errors. Instead, we used a truncated design consisting of two ground planes of 20 μm x 20 μm connected by vias around their perimeter and two simplest striplines: one straight-line and another one with a single 90-degree bend as shown in the Inset in Fig. 25. With the setting used, the total number of filaments was about 260 thousands, and the simulation time (on a PC with Intel@ Core i7-7660U CPU and 32 GB RAM) was about 11 minutes per each truncated layout. Using smaller maximum segment size to increase the accuracy (larger number of filaments) was not possible because this required much larger computer memory and computation time. The obtained results are shown in Fig. 26 and Fig. 27 by blue open circles (⊙). InductEx significantly overestimates inductance of the bends (corner inductance) as well as linear inductance (see [29]) as a result of a compromise between the accuracy and computation time.

We see that, at very large values $w \gtrsim 5$ μm ($w/d_1 \gtrsim 8$), the extracted $L_c/L_{sq}$ ratios indeed approach 0.56. However, they are noticeably, up to 30%, larger than the measured values at smaller $w$. At $w < 0.1$ μm, the numerically extracted corner inductance becomes negative. In order to check that this is not an artifact, we extracted corner inductance for other types of inductors in the SFQ5ee process: microstrips M6aM4 ($d_1 = 615$ nm), microstrips M6bM7 (identical to microstrips M1aM0, M2aM1, M3aM2, and M4aM3, etc., with $d_1 = 200$ nm, and symmetrical striplines M1aM0bM2 (identical to M2aM1bM3 and M3aM2bM4) with $d_1 = 200$ nm and $H = 600$ nm. The results are shown in Fig. 27.

The extracted inductance of right-angled bends of both types of the microstrips gives negative values of the corner inductance at $w/d_1 \lesssim 1$ in agreement with the literature data, e.g. [80,81], indicating that this is not an artifact caused by an insufficient accuracy of the extraction. However, at these small values, the results are clearly sensitive to the software settings and design details. For instance, two sets of points (+) for microstrips with $d_1 = 200$ nm correspond to the two sets of the extractions differing only by the number of grounding vias in the microstrip layouts.

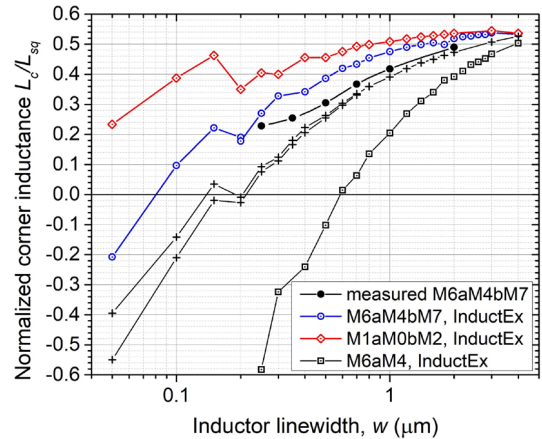

Fig. 27. Extracted inductance of right-angled bend (corner inductance) of various stripline and microstrip inductors in the SFQ5ee process: , from top to bottom: (◊) – striplines M1aMabM2 ($d_1 = 200$ nm, $H = 600$ nm); (⊙)- striplines M6aM4bM7; (●) – measured values for M6aM4bM7; (+) – microstrips M1aM0; (□) – microstrips M6aM4 ($d_1 = 615$ nm). Corner inductance is normalized to inductance per square $L_{sq} \equiv L_l w$. The ratio $L_c/L_{sq}$ asymptotically approaches 0.56, the value following from the analogy with the resistance of the right-angled bend of thin film conductors [86] only at $w > 5$ μm. We believe that negative values of $L_c/L_{sq}$ for the striplines at very small $w$ are likely an artifact of the insufficient accuracy. It is a real effect for all microstrips, $L_c/L_{sq}$ is negative at linewidths smaller than the dielectric thickness $w \lesssim d_1$.

*7.1.4 Discussion of the negative $L_c$.* Although negative corner inductance seems unphysical, it is simply a result of the definition (51), which takes the linear inductance equal to the $L_l$ of the straight-line inductor all the way up to the corner square. In reality, disturbance of the currents and magnetic fields caused by the bend is not confined to the corner square and spreads much farther out, reducing the linear inductance of the straight segments farther away from the corner square. Then negative $L_c$, purely mathematically assigned to a localized inductance of the corner, simply compensates for the overestimation of the distributed inductance of the straight segments near the corner.

A simple interpretation of the corner inductance can be given as related to a negative mutual inductance between the straight segments near the corner $M_b$. In this case we can present inductance of the bended wire as $L = L_l(l_1 + l_2 + w) - M_b$, from which $L_c = L_{sq} - M_b$. This mutual coupling exists only near the corner where two segments come very close to each other because mutual inductance of striplines exponentially decays with distance on the scale $p_0 = (H + 2\lambda)/\pi \approx 0.38$ μm for this type of parallel inductors. Theoretically, with decreasing $w$, the mutual inductance of the bend region may become larger than $L_{sq}$, resulting in the negative $L_c$. The maximum value of $M_b$ can be estimated from (40)-(41) as $M_l(p_0)p_0 \approx 0.046$ pH, where $M_l(p_0) \approx 0.12$ pH/μm (Fig. 14) is the linear mutual inductance of two filaments spaced at $p_0$. For M6aM4bM7 striplines, $L_{sq}$ becomes less than 0.04 pH only at $w < 0.02$ μm. Hence, negative $L_c$ cannot take place for M6aM4bM7 and other types of striplines at any practical linewidths, and negative $L_c$ extracted using InductEx should be viewed as an artifact caused by an insufficient accuracy of the extractor.

However, mutual inductance of microstrips is much larger than of the striplines and decreases with distance only as a power law $p^{-2}$, see eq. (19). Therefore, transition to negative corner inductance $L_c$ is possible at linewidths smaller than the dielectric thickness $w \lesssim d_1$, or below 0.6 μm for M6aM4 and below 0.2 μm for M1aM0 and similar types of the microstrips.

*7.2 Inductance of dense stripline serpentines*

Dense serpentines (meanders) use much smaller spacing between the adjacent segments than the sparse meanders, typically equal to the minimum allowed spacing; see Fig. 25c. As a results, at small linewidths, mutual coupling of many neighbors need to be taken into account to get an accurate inductance of the serpentine. To save space, we consider only the extreme case of $s = w = 0.25$ μm, the smallest allowed spacing in all of the MIT LL fabrication processes.

Electric current in the vertical segments $L_{y1}, L_{y2}, \ldots, L_{yN}$ of the meandered inductor in Fig. 25 flows in the alternating directions while it flows in the same direction in all horizontal segments. Therefore, any two adjacent vertical segments $i$ and $i \pm 1$ act as opposing inductors, while the next to the near neighbors $i$ and $i \pm 2$ act as adding inductors, and so on. For simplicity, we consider long meanders with length of the vertical segments $\Delta y \gg s, w$, so that $L_{y1} \gg L_{xi}$, where $L_{xi}$ is inductance of the horizontal segments. Extension to a more general case is trivial. At $\Delta y \gg s, w$, we can neglect coupling between $L_{xi}$ and $L_{xi+1}$ inductors on the opposite sides of the vertical inductor $L_{yi}$ because they are far away from each other, and neglect coupling between horizontal inductors on the same side. With these assumptions, inductance of the first L-section of the meander in Fig. 25(c) can be presented as

$$L_1 = L_{y1} - M_{12} + M_{13} - M_{14} + L_{x1}, \quad (60)$$

where plus and minus signs correspond to adding and opposing inductors, respectively. We included the corner inductance into the inductance of the vertical segment and neglected mutual coupling to all more distant neighbors because the distance to them is more than $3(s + w) \geq 1.5$ μm, so their mutual inductance, which decays exponentially with distance,

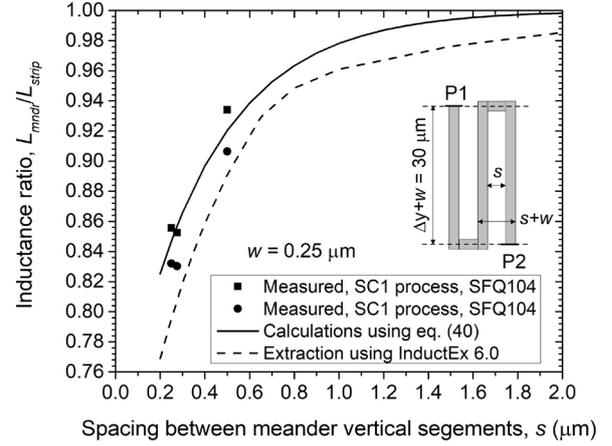

Fig. 28. Inductance of serpentine inductors M6aM4bM7 with three vertical sections as a function of spacing between them. Inductance was normalized to the measured inductance of a straight M6aM4bM7 stripline of the same length as the serpentine, measured along its central line. Data points (●) and (■) are the measured (on two different chips on the same wafer) values for M6aM4bM7 serpentines with $w = 0.25$ μm fabricated in the SC1 process, run SFQ104. Solid curve shows calculations using eq. (40) for the mutual inductance between the vertical segments. Dash curve shows the results of serpentine inductance extraction using InductEx 6.0 with the parameters of the SC1 process. Inset: sketch of the serpentine design. Parameters of the SC1 process used: $d_1 = 655$ nm, $H = 1.055$ μm, $\lambda = 90$ nm, and 200 nm for the thickness of all superconductor layers.

contributes much less than 1%. Similarly, the *i*-th L-section has inductance

$$L_i = L_{yi} - M_{i,i+1} - M_{i,i-1} + M_{i,i+2} + M_{i,i-2} - M_{i,i+3} - M_{i-3,i} + \cdots + L_{xi}, \quad (61)$$

where (...) stands for mutual inductance with more distant neighbors if account for them is needed. The total inductance of the meander is the sum of inductances of all L-sections.

For a uniform meander, inductances of all vertical sections are equal $L_{yi} = L_y$; equal are all horizontal inductances $L_{xi} = L_x$. Similarly, mutual inductances of all near neighbors are equal $M_{i,i+1} = M_{i-1,i} = M_{12}$; mutual inductances of the next to near neighbors are also equal $M_{i,i+2} = M_{i-2,i} = M_{13}$, if such neighbors exist, and so on. Also, the mutual inductance matrix is symmetrical. Hence, the total inductance between ports P1 and P2 in Fig. 25(c) becomes

$$L_{mndr} = NL_y - 2(N-1)M_{12} + 2(N-2)M_{13} - 2(N-3)M_{14} + \cdots + (N-1)L_x, \quad (62)$$

where $N$ is the number of the vertical segments.

The sum of the first and the last term in the (62) is the inductance of the straight line inductor $L_{strip} = L_l l$ with the length equal to the total length of the serpentine $l = N(\Delta y + w) + (N-1)s$. We neglected the corner inductance because it is very small at small linewidths (Fig. 26). The mutual inductance between two vertical segments $M_{ij} = M_l(\Delta y + 

$w$), where the linear mutual inductance $M_l$ can be calculated using expression (40) with distance $p_x = (|j - i|)(w + s)$ between them.

As an example, we measured M6aM4bM7 serpentines with $N = 3$ and $\Delta y + w = 30$ μm. The ratio of inductance of these serpentines to the inductance of the straight stripline of the same length and width can be expressed as

$$L_{mndr}/L_{strip} = 1 - \frac{4(\Delta y + w)}{l} M_l(\text{at } p_x = w + s) + \frac{2(\Delta y + w)}{l} M_l(\text{at } p_x = 2(w + s)). \quad (63)$$

$L_{strip}$ was measured independently on the same chip; it can be also calculated using eq. (45) or taken from Fig. 13. Mutual inductances $M_l$ at $p_x = w + s$ and at $p_x = 2(w + s)$ were calculated using (40) and also measured (see also Fig. 14 and Fig. 18). The measured inductance of $N = 3$ serpentines at a few values of $s$ is compared with expression (63) as well as with the results of InductEx extraction from the serpentine layouts in Fig. 28. Overall, there is a good, within 2.5% percent, agreement between the measured and the calculated using (63) inductance ratios, whereas InductEx gives somewhat lower (up to 6%) ratios.

## 7. Conclusion

We studied mutual and self-inductance of all typical inductors encountered in advanced multilayered superconductor integrated circuits using planarized layers and mostly sub-micrometer linewidths comparable to the thickness of the superconductor layers and interlayer dielectrics. We presented a set of simple analytical expressions describing: self-inductance of superconducting microstrips, exp. (21), and striplines, exp. (37) and (45); mutual inductance of superconducting microstrips, exp. (19) and (26), and striplines, exp. (39) and (40). Their accuracy is better than 2% in entire the range of linewidths and thicknesses of interest to the very large scale integrated (VLSI) circuits, typically $0 < w \leq 2$ μm. These expressions can be used in computer-assisted design (CAD) tools to automatically calculate inductance while inductors are being drawn as well as for calibration of the numerical inductance extractors.

We demonstrated that mutual inductance of stripline inductors (two ground planes) decays exponentially with distance between the strips with the decay length $(H + 2\lambda)/\pi$, whereas mutual inductance of microstrip inductors (one ground plan) is long-ranged and, at large distances, decreases only as $(h_1 + \lambda)^2/(s + w)^2$ with spacing between them. Therefore, in the fabrication processes developed at MIT LL (the SFQ5ee, the SC1, and the SC2 processes), mutual coupling between any striplines can be neglected if spacing between them $s \gtrsim 1$ μm, whereas coupling between the microstrips remains strong even at $s \sim 4$ μm. Therefore, stripline inductors should be exclusively used in VLSI circuits to minimize parasitic coupling between adjacent inductors and transformers. The use of microstrip-based inductors and transformers is not recommended in dense circuits due to a large and poorly controlled parasitic coupling between them.

Perforations in the ground plane (long slits placed along the inductors) have relatively small effect on the self-inductance but substantially increase mutual inductance of inductors near or within the slit, a few times with slits of moderate width ~2 μm. Long slits with sparse narrow bridges across them have nearly the same effect as slits without bridges.

We also measured inductance of right-angled bends of striplines and found that, contrary to the general belief, corner inductance strongly decreases with decreasing the linewidth and becomes negligible below $w \approx 0.25$ μm. At very large linewidth $w > 5$ μm, the corner inductance asymptotically approaches $0.56 L_{sq}$, the value following from analogy between inductance of the bend and the bend contribution to the resistance of thin film resistors. The corner inductance of microstrips becomes negative at linewidths smaller than the dielectric thickness between the strip and the ground plane.

A surprising observations for us was that mutual inductance of superconducting inductors in the range of linewidths and thicknesses relevant to integrated circuits does not depend on superconductivity of their signal traces and only depends on the magnetic field penetration depth in superconducting ground planes and the geometrical dimensions. The normal metal signal traces with superconducting ground planes would have the same mutual inductance as superconducting traces. The main source of variability (parameter spreads) of superconducting transformers is the exponential dependence of mutual inductance on spacing between the parallel traces and a very strong dependence on dielectric thickness between them, whereas dependence on the linewidth is weak.

We compared the experimental data and the analytical expressions with numerical simulations using wxLC software of M. Khapaev [54-56] and numerical extraction results using InductEx 6.0 developed by C. Fourie [60]. For uniform transmission line structures, wxLC provides superior accuracy and better agreement with the presented analytical expressions and the experimental data. However, for truly 3D structures InducEx is indispensable tool allowing for mutual and self-inductance extraction from the actual circuit layout with reasonable accuracy and in a reasonable amount of time.

The presented set of data is sufficient for designing superconductor integrated circuits in the fabrication processes available at MIT Lincoln Laboratory such as the SFQ5ee and the SC1/SC2 processes, whereas the presented analytical expressions can be used for calculating mutual and self-inductance of inductors in any fabrication process for superconductor electronics or with normal metals.

And the last but not the least, it follows from the results that superconducting transformers are poorly scalable component of superconductor integrated circuits due to a weak dependence of mutual inductance of vertically or horizontally spaced inductors on the linewidth and, as a consequence, very weak dependence of the transformer area on the linewidths. This is because self-inductance of the transformer's primary and secondary strongly increases with decreasing the linewidth due to the growing as $w^{-1}$ kinetic inductance, which does not contribute to mutual coupling, while the current carrying capacity of the wires decreases linearly with decreasing $w$. These issues and their consequences for scalability of superconductor electronics using ac powering and clocking schemes will be discussed in detail elsewhere.

## Acknowledgements

We are grateful to Vasili Semenov for numerous discussions of inductance measurements and extraction. S.K. Tolpygo would like to thank Mikhail M. Khapaev for the access to inductance extraction software wxLC, and to Coenrad J. Fourie for the access to and help with InductEx 6.0. The authors are grateful to Alex Wynn for developing and maintaining a database of inductance measurements, to Ravi Rastogi and Scott Zarr for their help with wafer fabrication, and to the entire fabrication team of the Microelectronics Lab at MIT Lincoln Laboratory. We would like to thank Leonard Johnson, Mark Gouker, Scott Holmes, and Mark Heiligman for their interest in and support of this work.

This research was based upon work supported in part by the Office of the Director of National Intelligence (ODNI), Intelligence Advanced Research Projects Activity (IARPA), via Air Force Contract FA8702-15-D-0001, and in part by the Under Secretary of Defense for Research and Engineering under Air Force Contract No. FA8702-15-D-0001. Any opinions, findings, conclusions or recommendations expressed in this material are those of the authors and do not necessarily reflect the views of the Under Secretary of Defense for Research and Engineering and should not be interpreted as necessarily representing the official policies or endorsements, either expressed or implied, of the ODNI, IARPA, or the U.S. Government.